\documentclass[iop,revtex4]{emulateapj}
\shorttitle{Optical spectroscopy of redbacks}
\shortauthors{Strader \etal~}
\def\etal{{et al.}}

\def\hub{\ifmmode H_\circ\else H$_\circ$\fi}
\usepackage{subfigure}
\usepackage{amsmath}
\usepackage{graphics}
\usepackage{hyperref}
\usepackage{breakurl} 
\usepackage{pdflscape}

\def\ltsima{$\; \buildrel < \over \sim \;$}
\def\simlt{\lower.5ex\hbox{\ltsima}} 
\def\gtsima{$\; \buildrel > \over \sim \;$}
\def\simgt{\lower.5ex\hbox{\gtsima}} 
\def\arcsec{\hbox{$^{\prime\prime}$}}

\begin{document}

\title{Optical spectroscopy and demographics of redback millisecond pulsar binaries}
\author{
Jay Strader\altaffilmark{1},
Samuel J.~Swihart\altaffilmark{1},
Laura Chomiuk\altaffilmark{1}, 
Arash Bahramian\altaffilmark{2},
Christopher T.~Britt\altaffilmark{3},
C.~C.~Cheung\altaffilmark{4},
Kristen C.~Dage\altaffilmark{1},
Jules P.~Halpern\altaffilmark{5}, 
Kwan-Lok Li\altaffilmark{6},
Roberto P.~Mignani\altaffilmark{7,8}, 
Jerome A.~Orosz\altaffilmark{9}, 
Mark Peacock\altaffilmark{1},
Ricardo Salinas\altaffilmark{10},
Laura Shishkovsky\altaffilmark{1},
Evangelia Tremou\altaffilmark{11}
}
\altaffiltext{1}{Center for Data Intensive and Time Domain Astronomy, Department of Physics and Astronomy, Michigan State University, East Lansing, MI 48824, USA; strader@pa.msu.edu}
\altaffiltext{2}{International Centre for Radio Astronomy Research, Curtin University, GPO Box U1987, Perth, WA 6845, Australia}
\altaffiltext{3}{Space Telescope Science Institute, 3700 San Martin Drive, Baltimore, MD, 21218, USA}
\altaffiltext{4}{Space Science Division, Naval Research Laboratory, Washington, DC, 20375, USA}
\altaffiltext{5}{Department of Astronomy, Columbia University, 550 West 120th Street, New York, NY 10027, USA}
\altaffiltext{6}{Department of Physics, Ulsan National Institute of Science and Technology, Ulsan 44919, Republic of Korea}
\altaffiltext{7}{INAF - Istituto di Astrofisica Spaziale e Fisica Cosmica Milano, via E. Bassini 15, 20133, Milano, Italy}
\altaffiltext{8}{Janusz Gil Institute of Astronomy, University of Zielona G\'ora, ul Szafrana 2, 65-265, Zielona G\'ora, Poland}
\altaffiltext{9}{Department of Astronomy, San Diego State University, 5500 Campanile Drive, San Diego, CA 92182, USA}
\altaffiltext{10}{Gemini Observatory, Casilla 603, La Serena, Chile}
\altaffiltext{11}{AIM, CEA, CNRS, Universit{\'e} Paris Diderot, Sorbonne Paris Cit{\'e},  Universit{\'e} Paris-Saclay, F-91191 Gif-sur-Yvette, France.}

\begin{abstract}

We present the first optical spectroscopy of five confirmed (or strong candidate) redback millisecond pulsar binaries, obtaining complete radial velocity curves for each companion star. The properties of these millisecond pulsar binaries with low-mass, hydrogen-rich companions are discussed in the context of the 14 confirmed and 10 candidate field redbacks. We find that the neutron stars in redbacks have a median mass of $1.78\pm0.09 M_{\odot}$ with a dispersion of $\sigma = 0.21\pm0.09$. Neutron stars with masses in excess of $2 M_{\odot}$ are consistent with, but not firmly demanded by, current observations. Redback companions have median masses of $0.36\pm0.04 M_{\odot}$ with a scatter of $\sigma = 0.15\pm0.04 M_{\odot}$, and a tail possibly extending up to 0.7--$0.9 M_{\odot}$. Candidate redbacks tend to have higher companion masses than confirmed redbacks, suggesting a possible selection bias against the detection of radio pulsations in these more massive candidate systems. The distribution of companion masses between redbacks and the less massive black widows continues to be strongly bimodal, which is an important constraint on evolutionary models for these systems. Among redbacks, the median efficiency of converting the pulsar spindown energy to $\gamma$-ray luminosity is $\sim 10\%$.

\end{abstract}
 
\keywords{X-rays: binaries --- binaries: spectroscopic --- pulsars: general --- stars: neutron}

\section{Introduction}

Redbacks are millisecond pulsar binaries with low-mass ($\lesssim 1 M_{\odot}$), hydrogen-rich companions (Roberts 2013). They are typically distinguished from the even lower mass black widows (with $M_c \lesssim 0.05 M_{\odot}$) by having companion masses $\gtrsim 0.1 M_{\odot}$. Redbacks provide important tests of the physics of neutron star formation and the recycling of millisecond pulsars. They are often difficult to detect as radio pulsars with timing observations, due to eclipses associated with material from the companions. Their discovery in substantial numbers has been facilitated by the \emph{Fermi} Large Area Telescope (LAT), since the $\gamma$-ray emission, thought to come from the millisecond pulsar, is not affected by similar selection biases (e.g., Ray et al.~2012). Two field redbacks (PSR J1023+0038 and XSS J12270--4859) also belong to the unusual class of transitional millisecond pulsars, switching between distinct states of accretion-powered and rotation-powered emission on timescales of weeks to years (Archibald et al.~2009; Bassa et al.~2014; Roy et al.~2015).

Optical observations of individual redbacks provide important constraints on the component masses, inclination, and distance of each binary. This paper presents optical spectroscopy of four redbacks confirmed as pulsars: PSR J1048+2339 (Deneva et al.~2016; Cromartie et al.~2016), PSR J1431--4715 (Bates et al.~2015), PSR J1622--0315 (Sanpa-Arsa 2016), PSR J1628--3205 (Ray et al.~2012; Li et al.~2014). We also include one candidate redback, 3FGL J2039.6--5618 (Salvetti et al.~2015). In the latter case we show the optical binary likely contains a neutron star and hence is almost certainly associated with the $\gamma$-ray source, making it a probable redback. 

We then discuss the properties of these five systems in the context of other redbacks.
As of the end of 2018, 24 systems in the field are confirmed or good candidates for being redbacks (see \S 4 for a compilation of these sources).
Of these, 18 (75\% of the total) were discovered through follow-up of previously unassociated \emph{Fermi}-LAT GeV $\gamma$-ray sources.
Nearly half of the redback discoveries (11 of 24) have been published since 2016, showing that the pace of discovery is not slowing.

\section{Data}

For this study, we selected confirmed or candidate redback systems that did not yet have optical spectroscopy and which were observable from the SOAR telescope in Chile. 

Additional data were also obtained for other systems and are being presented in full elsewhere (PSR J1306--40, Swihart et al., in preparation; 3FGL J0954.8--3948, Li et al.~2018), though their basic orbital properties are used in \S 4 below.

The spectroscopic observations were all made using the Goodman Spectrograph (Clemens et al.~2004) on the SOAR telescope. Good spectra were obtained for a minimum of 13 epochs (PSR J1048+2339) to a maximum of 44 epochs (3FGL J2039.6--5618). Individual exposure times ranged from 10--25 min per spectrum, depending on the prevailing conditions and the brightness of the optical source (which ranged from $V \sim 17.5$ to $V > 20$ mag). Most observations were made with a 400 l mm$^{-1}$ grating and a 0.95\arcsec\ slit, yielding a full-width at half maximum (FWHM) resolution around 5.6 \AA. Observations prior to late 2016 all used the Goodman ``blue" camera, with a typical wavelength coverage of $\sim 3400$--7000 \AA\ using the 400 l mm$^{-1}$ grating. Most data since that time used the Goodman ``red" camera, with typical wavelength coverage of  $\sim 3800$--7800 \AA\ (for 3FGL J2039--5618, this was $\sim 4800$--8800 \AA). Owing to its higher optical flux, for PSR J1431--4715 we were able to use a 1200 l mm$^{-1}$ grating, giving a resolution of about 2.1 \AA\ over 5500--6730 \AA. Wavelength calibration was accomplished using FeAr arcs taken every 30--60 min during the night, with small zeropoint corrections made using night sky lines.

All spectra were reduced and optimally extracted in the standard way. Radial velocities were determined by cross-correlation with bright standards taken with the same setup. The region used for the cross-correlation varied by target. For PSR J1622--0315, PSR J1628--3205, and 3FGL J2039.6--5618, we mostly used the region around the Mg$b$ triplet, though for some spectra with lower signal in the bluer region, we used the feature-rich region from 6050 to 6250 \AA. The optical counterpart of PSR J1048+2339 is cooler, with broad molecular bands, so we used a wider wavelength region for cross-correlation. The counterpart of PSR J1431--4715 has weak metal lines and so we used the strong H$\alpha$ absorption line; for a few spectra where metal-line velocities were also measurable, the metal-line and Balmer velocities were consistent. Additional discussion of the spectra themselves can be found in \S 3.6. 

All velocities were corrected to the Solar System barycenter and associated with Modified Barycentric Julian Dates (MBJD, BJD -- 2400000.5) on the TDB system (Eastman et al.~2010). The velocities are given in Tables 1--5.

\section{Results and Analysis}

We fit circular Keplerian models to the radial velocities derived in \S 2 using {\tt rvfit} (Iglesias-Marzoa et al.~2015). For each system, excepting 3FGL J2039.6--5618 (see \S 3.5), significant eccentricity is rejected by the pulsar observations or not preferred by the optical data. In the absence of additional constraints, the circular model has four free parameters: period $P$, epoch of the ascending node $T_0$, secondary semi-amplitude $K_{2}$, and systemic velocity $\gamma$. For the confirmed pulsars a subset of these parameters (typically $P$ and often $T_0$) are known, leaving fewer free parameters than four. In all cases the uncertainties in the fits were derived through bootstrapping. 

The measurement of $P$ and $K_2$ immediately allows the calculation of the mass function $f(M) = P K_2^3/(2 \pi G) = M_1 (\textrm{sin} \, i)^3/(1+q)^2$ for gravitational constant $G$, inclination angle $i$, and mass ratio $q = M_1/M_2 = M_c/M_{NS}$. For systems with pulsar timing, the combination of the pulsar and companion mass functions give a direct measurement of $q$. The final and most uncertain quantity needed to  solve the system is the inclination, for which constraints of variable accuracy and precision can be available from optical light curve fitting. Light curve data have already been published for some of our systems, and we present new data or analysis for PSR J1431--4715 and 3FGL J2039.6--5618.

In the following discussion, when no other constraints are available, we assume the neutron star primary is in the mass range 1.4--2.0 $M_{\odot}$, consistent with most known millisecond pulsars with precise measurements (e.g., Lattimer 2012;  {\"O}zel \& Freire 2016; Antoniadis et al.~2016). Some recycled neutron stars might fall slightly below (e.g., Fonseca et al.~2016) or above (e.g., van Kerkwijk et al.~2011; Linares et al.~2018) this range, but this would not change any of the conclusions for which this assumption is relevant. Additional discussion about the redback neutron star mass distribution can be found in \S 4.3.

Figures 1--5 show the circular Keplerian fits to the SOAR/Goodman barycentric radial velocities listed in Tables 1--5 for each redback. A description of the results presented in these figures can be found in the respective subsections below.

\subsection{PSR J1048+2339}

\begin{figure}[t]
\includegraphics[width=3.4in]{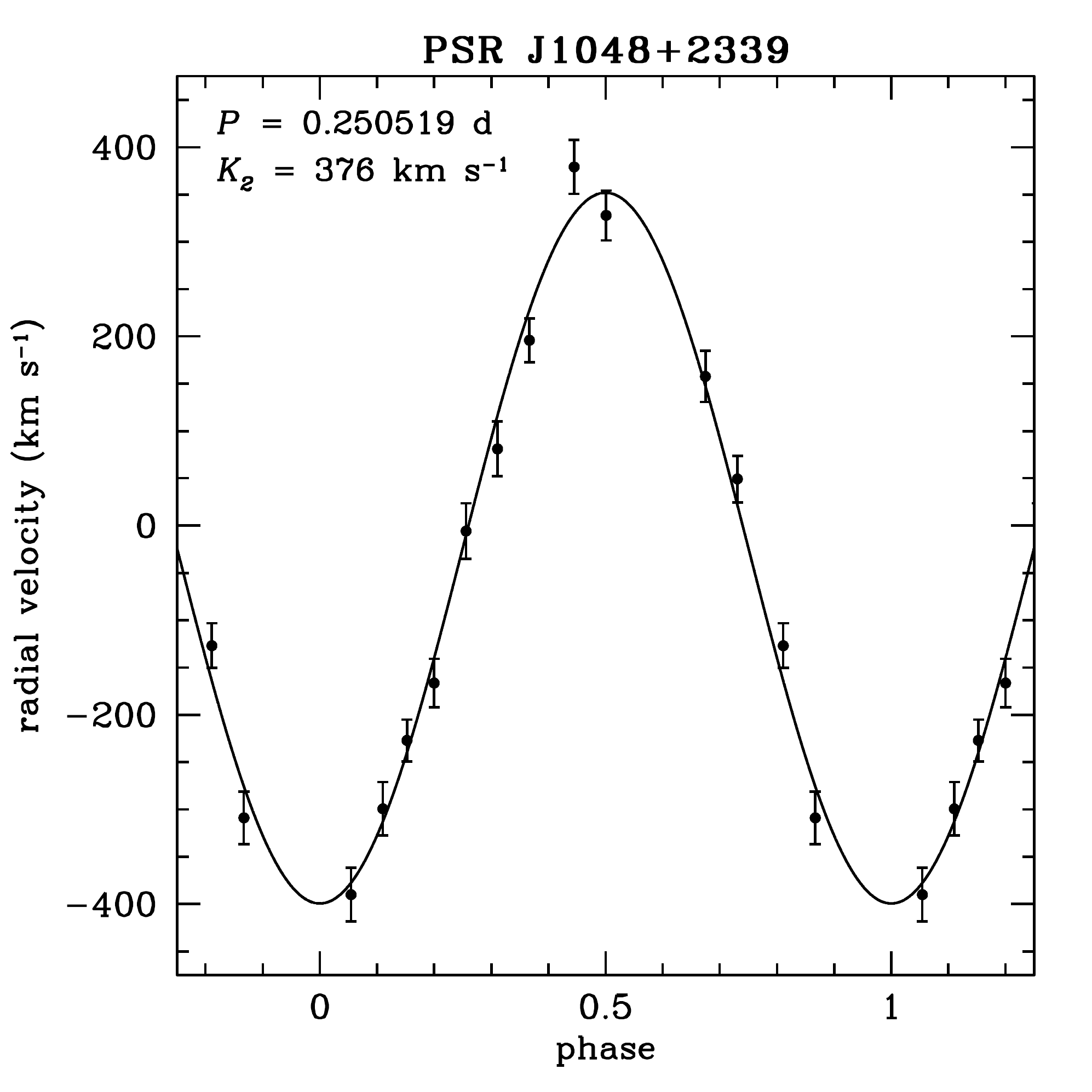}
\caption{Circular Keplerian fit to the SOAR/Goodman barycentric radial velocities of PSR J1048+2339.}
\label{fig:1048_rv}
\end{figure}

This pulsar was discovered in an Arecibo search of unassociated \emph{Fermi}-LAT sources by Cromartie et al.~(2016). Since the pulsar ephemerides for this source have been published by Deneva et al.~(2016), the only free parameters in our fit are $K_2$ and $\gamma$. We find $K_2 = 376(14)$ km s$^{-1}$ and $\gamma = -24(8)$ km s$^{-1}$, where the numbers in parenthesis represent the $1\sigma$ uncertainty in the last 1 or 2 digits.  A fit with these median values has a $\chi^2$/d.o.f. of 14/11 (Figure 1). The reduced $\chi^2$ is $> 1$ largely due to a single outlying measurement. The mass function $f(M) = 1.37(15) M_{\odot}$. Combined with the companion mass function from Deneva et al.~(2016), we find $q=0.194(7)$. Remarkably, this implies that the minimum mass of the neutron star is $M_{NS} = 1.96(22) M_{\odot}$, and hence the orbit must be very close to edge-on, unless the neutron star has a mass much larger than $2 M_{\odot}$. For  $M_{NS} = 1.96(22) M_{\odot}$, the companion mass $M_c = 0.38(4) M_{\odot}$ (note that all mass estimates are collected in Table 7).

{We note that if the single outlying velocity measurement is removed and the analysis repeated, then $K_2 = 363(11)$ km s$^{-1}$ and the minimum neutron star mass is $M_{NS} = 1.76(17) M_{\odot}$. Since these values are within the uncertainties of the original estimates, and there is no unambiguous justification for removing a velocity datum, we use the results of the initial analysis for the remainder of the paper. Additional radial velocity measurements of the companion to PSR J1048+2339 would be extremely valuable to improve the precision of the neutron star mass constraint.}

Despite the likely near edge-on orbit, there is no evidence for total X-ray eclipses in the \emph{Chandra} X-ray light curves presented by Cho et al.~(2018), although they do observe a broad minimum centered around $\phi = 0.25$ (when the companion is in front of the neutron star). This indicates that the X-rays are not produced only in the immediate vicinity of the pulsar, but are being emitted from a larger region that is never fully eclipsed along our line of sight. Cho et al.~(2018) suggest that the X-rays could be produced in a shock that wraps around the pulsar, {an idea also discussed in the literature for other black widows and redbacks (e.g., Romani \& Sanchez 2016; Wadiasingh et al.~2017).}

\subsection{PSR J1431--4715}

\begin{figure}[t]
\includegraphics[width=3.4in]{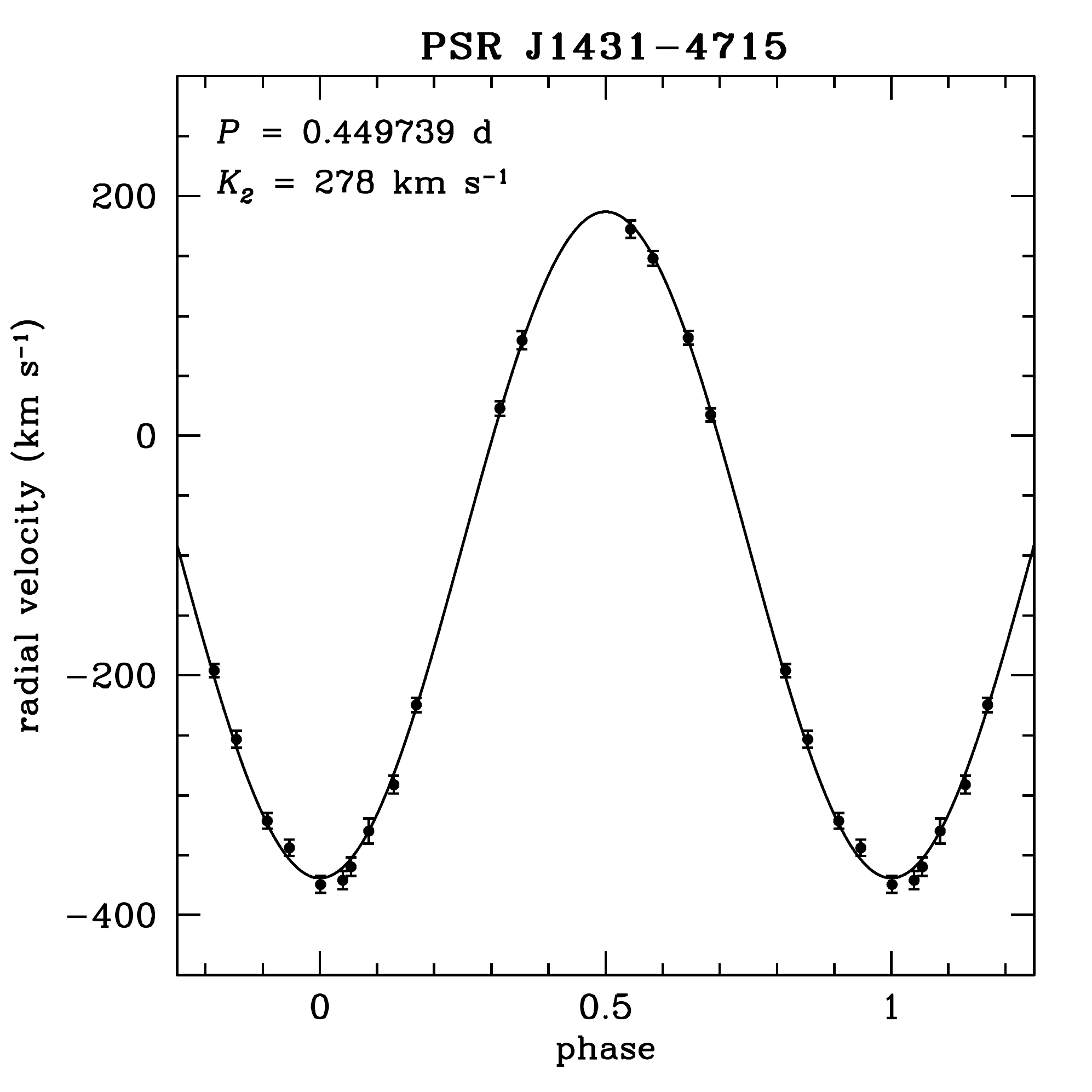}
\caption{Circular Keplerian fit to the SOAR/Goodman barycentric radial velocities of PSR J1431--4715.}
\label{fig:j1431_rv}
\end{figure}

PSR J1431--4715 was discovered by the High Time Resolution Universe Pulsar Survey (Bates et al.~2015) and as such is one of the few redbacks discovered independently of \emph{Fermi}; the source is not present in the 4-year 3FGL catalog (Acero et al.~2015), but has a likely association in the preliminary 8-year \emph{Fermi}-LAT catalog\footnote{https://fermi.gsfc.nasa.gov/ssc/data/access/lat/fl8y/}. We use $P$ and $T_0$ from the pulsar timing paper of Bates et al.~(2015) and fit only $K_2$ and $\gamma$. We find $K_2 = 278(3)$ km s$^{-1}$ and $\gamma = -91(2)$. The circular fit has a $\chi^2$/d.o.f. of 10/14 (Figure 2), with a low rms of 5.6 km s$^{-1}$, suggesting an excellent model fit. The optical spectra give $f(M) = 1.00(3) M_{\odot}$, and also $q = 0.096(1)$ when combined with the pulsar timing data. For a $1.4 M_{\odot}$ primary we find that  $i = 72(2)^{\circ}$, while a $2.0 M_{\odot}$ primary implies $i = 58(1)^{\circ}$. For this neutron star mass range, the companion mass lies in the range 0.13--$0.19 M_{\odot}$.

\subsubsection{Optical Light Curve Modeling}

\begin{figure}[t]
\hspace{-0.3cm}
\includegraphics[width=3.6in]{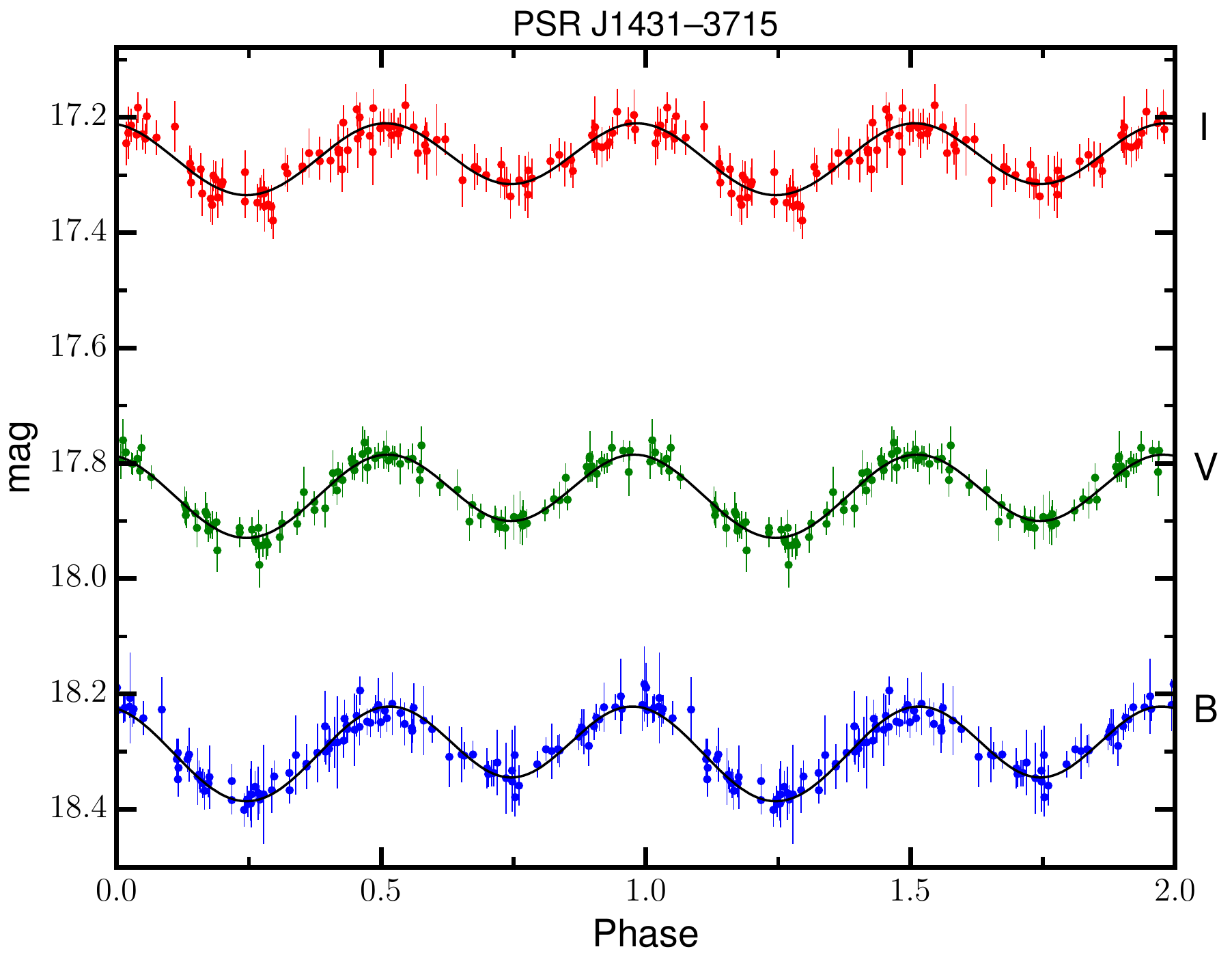}
\caption{{\tt ELC} model fits to the $BVI$ light curves of PSR J1431--4715, as described in the text.}
\label{fig:j1431_lc}
\end{figure}

We also obtained optical $BVI$ photometry of this system with ANDICAM on the SMARTS 1.3-m telescope at CTIO in early 2018. These data are given in Table 6. When folded on the known ephemerides, the most obvious feature of the these light curves is double peaked ellipsoidal variations, but modified such that the ``day'' side of the companion ($\phi$=0.75) is brighter than the ``night'' side, consistent with heating of the secondary from the pulsar or perhaps an intrabinary shock (e.g., Cho et al.~2018).

Using the pulsar ephemerides and our optical spectroscopic constraints, we modeled these data using {\tt ELC} (Orosz \& Hauschildt 2000) in the same way as for 3FGL J2039.6--5618 (see \S3.5.1). We assume a mean intensity-weighted secondary surface $T_{\rm eff}$ = 6500 K (consistent with spectroscopy) and fit for the inclination $i$, the Roche lobe filling factor $f_2$, a phase shift $\Delta \phi$, and level of irradiation heating. We assume foreground reddening from Schlafly \& Finkbeiner (2011) of $E(B-V) = 0.142$.

In all our best-fitting models, we find the  day side of the secondary is heated to an effective temperature $\sim$130--200 K hotter than the night side. Additionally, a small phase shift of $\Delta \phi \sim$ --0.005 (equivalent to $\sim 3.2$ min) is required to fit the light curves. This detail can be seen most clearly around $\phi$=1.0, where the light curves peak slightly prior to that expected from simple ellipsoidal modulations. Such offsets are not unusual among redbacks (see also \S 3.5.1).

Unfortunately, while the overall scale of the binary system is well-constrained, our light curve modeling suggests the star is far from filling its Roche lobe (with a filling factor from $\sim$ 0.64 to 0.76), and hence no useful constraints on the inclination or masses of the components are possible. For example, models that have $f_2 = 0.64$ and $i = 83^{\circ}$ (implying $M_{NS} = 1.23 M_{\odot}$) give fits of identical quality to models with $f_2 = 0.76$ and $i = 53^{\circ}$ (for which $M_{NS} = 2.33 M_{\odot}$). We show the high-inclination model fit to the data in Figure 3, but emphasize that model fits for essentially any reasonable neutron star mass would be of identical quality.

The companion to PSR J1431--4715 is bright enough that a precise \emph{Gaia} parallax measurement may be possible in a future \emph{Gaia} data release. This would give an independent constraint on the radius of the secondary and hence help break the degeneracy between inclination and filling factor.

\subsection{PSR J1622--0315}

Sanpa-Arsa (2016) discuss an initial timing solution of the redback PSR J1622--0315, which was discovered in a Green Bank Telescope search of \emph{Fermi}-LAT unassociated sources. We adopt their orbital period of 0.1617006798(6) d. For a circular model, we fit $T_0$, $K_2$, and $\gamma$. We find a best-fit model of $T_0 = 58160.3522(4)$ d (MBJD), $K_2 = 423(8)$ km s$^{-1}$, and $\gamma = -135(6)$  km s$^{-1}$. This fit has a $\chi^2$/d.o.f. of 15/11 (Figure 4) and an rms of 18 km s$^{-1}$. The orbital measurements give $f(M) = 1.27(7) M_{\odot}$, consistent with a relatively edge-on orientation. Bringing in the pulsar projected semi-major axis, we find $q = 0.070(1)$. 
This implies a minimum MSP mass of $1.45(8) M_{\odot}$ for an edge-on orbit. If instead we assume the MSP mass is  $2.0 M_{\odot}$, the 
 inclination $i = 64(2)^{\circ}$. The secondary has a relatively low mass for a redback, 0.10--0.14 $M_{\odot}$.

\begin{figure}[t]
\includegraphics[width=3.4in]{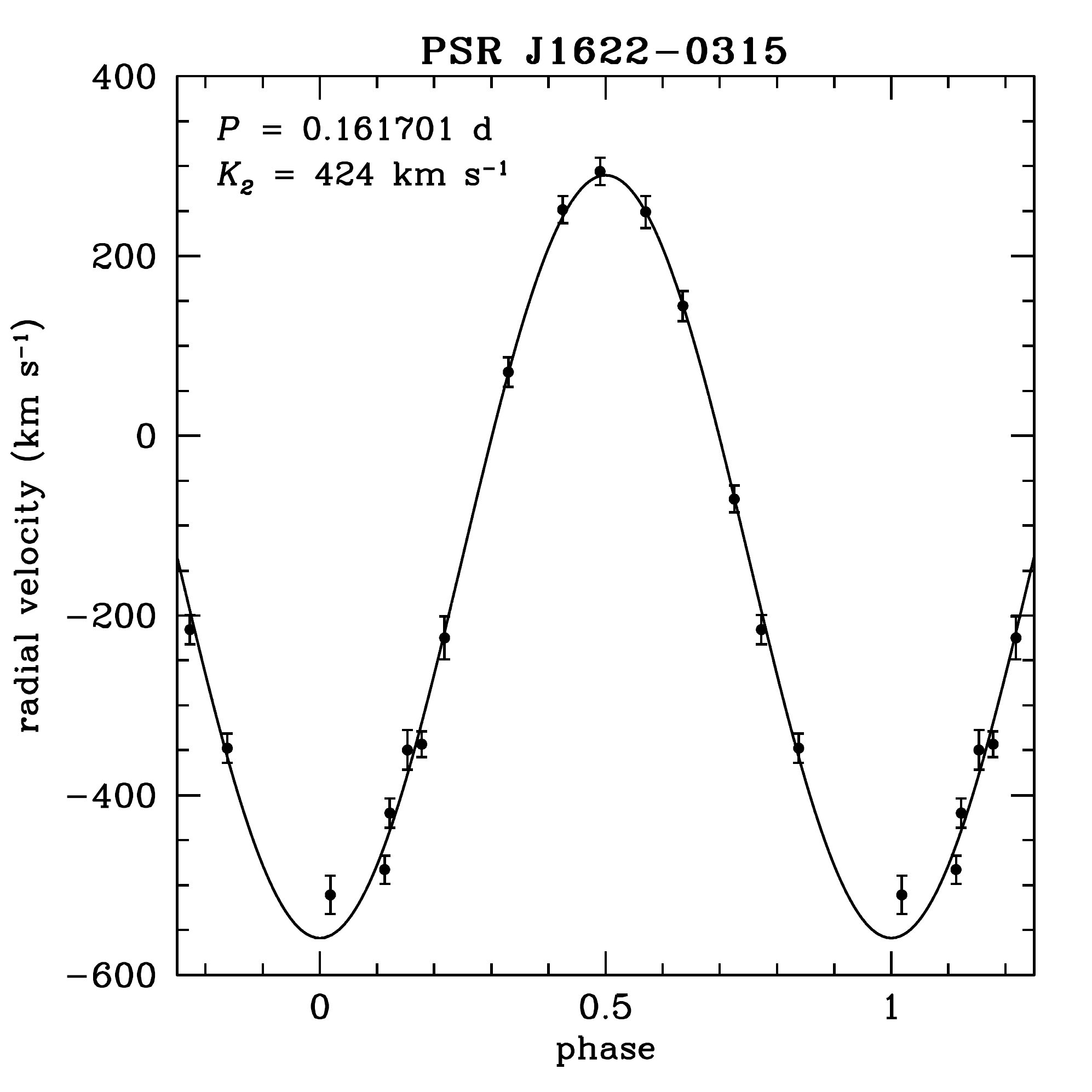}
\caption{Circular Keplerian fit to the SOAR/Goodman barycentric radial velocities of PSR J1622--0315.}
\label{fig:j1622_rv}
\end{figure}

\subsection{PSR J1628--3205}

PSR J1628--3205 was discovered by the Green Bank Telescope in a radio pulsation search of unassociated \emph{Fermi} $\gamma$-ray sources (Ray et al.~2012). Li et al.~(2014) identified the $V$ $\sim 20$ optical counterpart to the pulsar, and their photometry showed both ellipsoidal variations and evidence for variable heating.

For the orbital fit, $P$ and $T_0$ were fixed to the pulsar ephemerides (Cho et al.~2018), leaving only $K_2$ and $\gamma$ to be fit. We find $K_2 = 358(10)$ km s$^{-1}$ and $\gamma = -4(7)$ km s$^{-1}$. These values give a fit with an rms of 20 km s$^{-1}$ and a $\chi^2$/d.o.f. of 11/15 (Figure 5). The optical orbital elements imply  $f(M) = 0.99(8) M_{\odot}$. Since PSR J1628--3205 has its projected semi-major axis $a$ sin $i$ measured,  our $K_2$ value immediately gives the mass ratio $q = 0.120(3)$. For an assumed $M_{NS} = 1.4 M_{\odot}$, $i = 74(6)^{\circ}$, and for $M_{NS} = 2.0 M_{\odot}$, $i = 59(3)^{\circ}$.  For this range of neutron star masses, the secondary must be in the mass range 0.17--0.24 $M_{\odot}$. These inclination constraints are in the general range of the values inferred from the light curve fitting in Li et al.~(2014). Improved light curve fits could allow stronger constraints on the inclination and hence on the mass of the binary components.

\begin{figure}[t]
\includegraphics[width=3.4in]{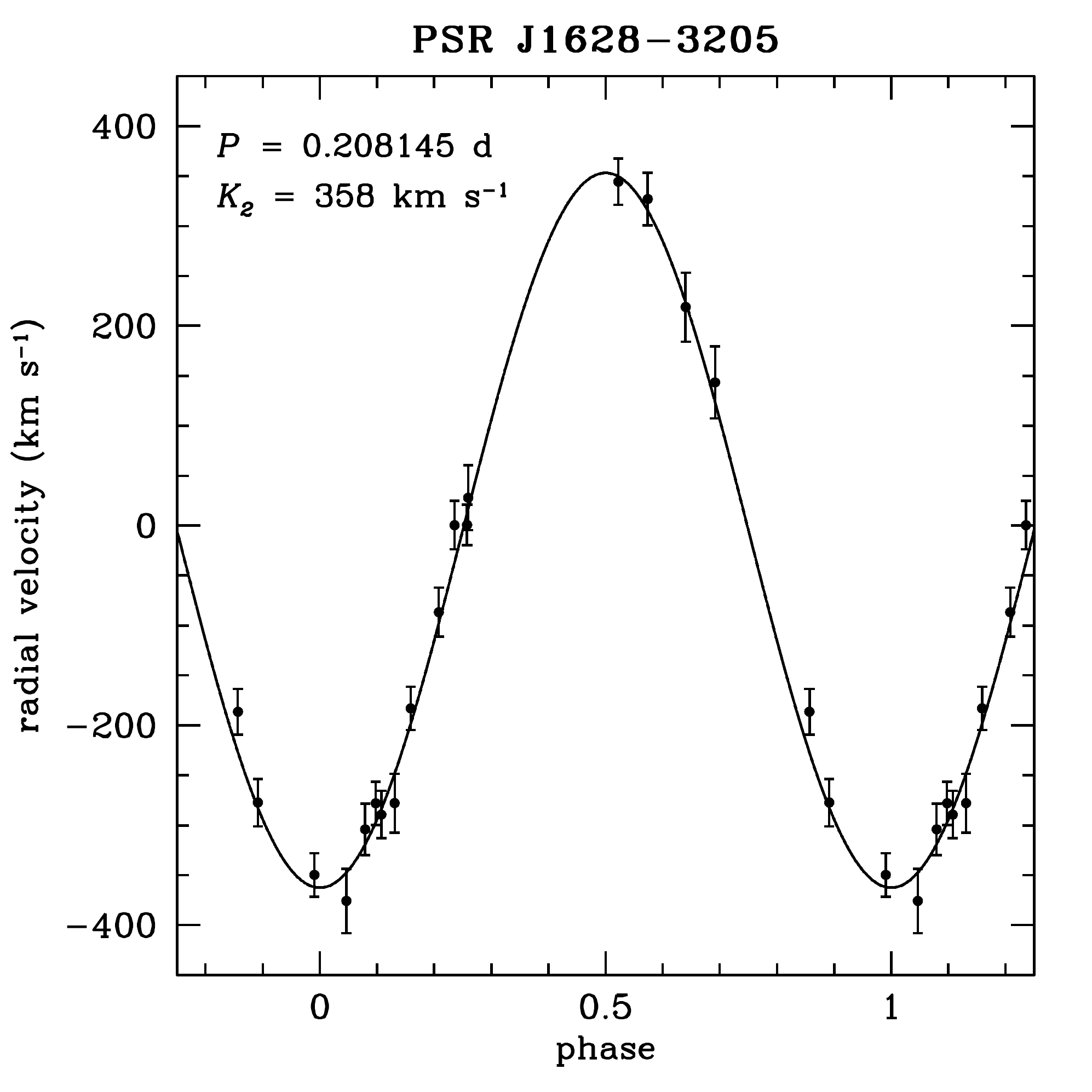}
\caption{Circular Keplerian fit to the SOAR/Goodman barycentric radial velocities of PSR J1628--3205.}
\label{fig:j1628_rv}
\end{figure}

\subsection{3FGL J2039.6--5618}

Since no pulsar has yet been detected in this candidate redback, it is worth discussing the background for this binary. Salvetti et al.~(2015) and Romani (2015) used X-ray and optical observations of the \emph{Fermi}-LAT error ellipse of the source to identify a likely counterpart: they found an object that was the brightest X-ray source in \emph{XMM-Newton} data covering the 3FGL \emph{Fermi} error ellipse. Furthermore, this source showed X-ray and optical variability with the same 0.23 d period, consistent with the expected properties of a compact binary. We note that in the preliminary 8-year \emph{Fermi}-LAT catalog\footnote{https://fermi.gsfc.nasa.gov/ssc/data/access/lat/fl8y/)}, soon to be superseded by the official 4FGL catalog, this X-ray source is close to the center (0.23\arcmin) of the $\sim 1.3\arcmin$ radius 95\% \emph{Fermi}-LAT error ellipse, consistent with the proposed association.

\begin{figure}[t]
\includegraphics[width=3.4in]{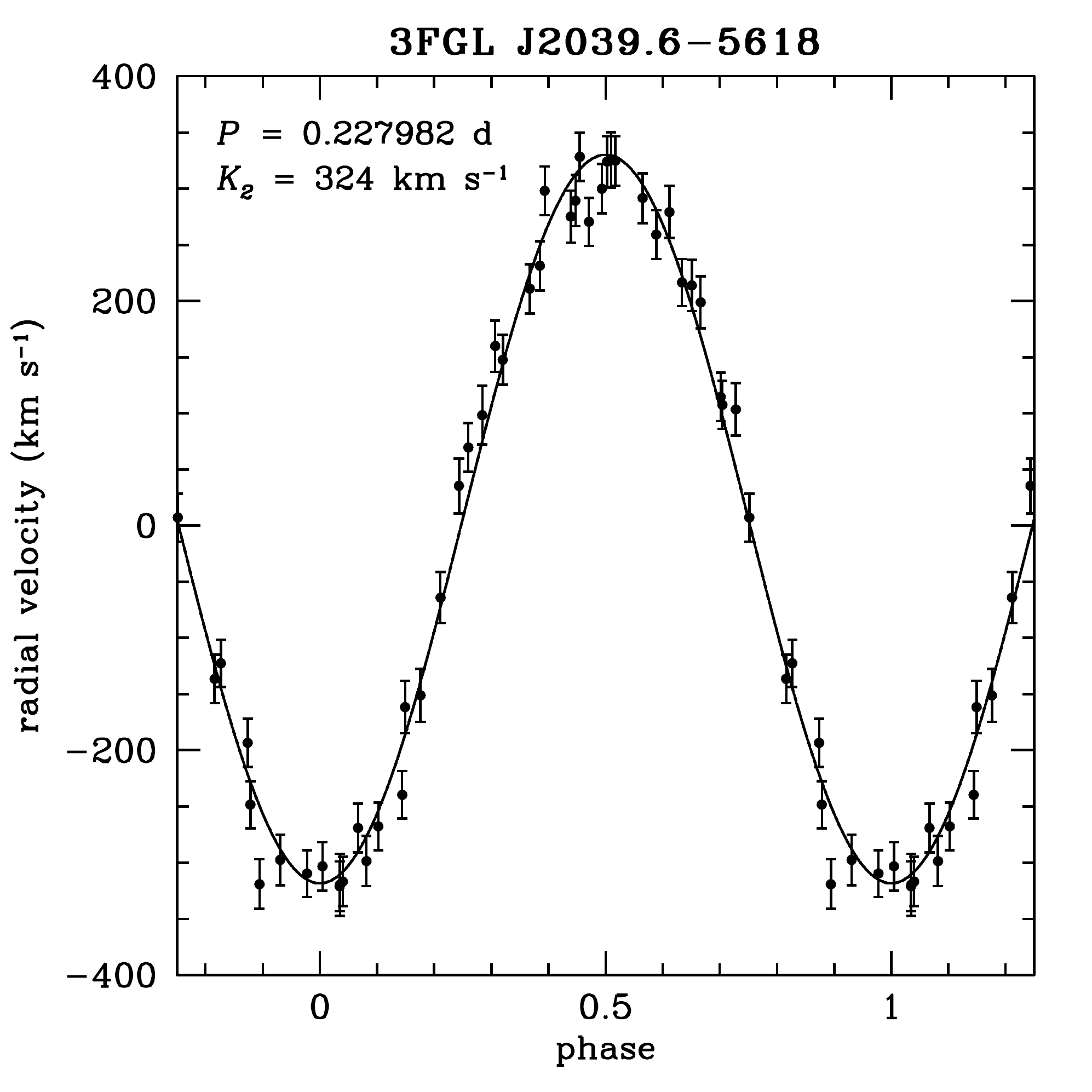}
\caption{Circular Keplerian fit to the SOAR/Goodman barycentric radial velocities of 3FGL J2039.6--5618.}
\label{fig:j2039_rv}
\end{figure}

Since no pulsar has yet been detected in this system, a circular Keplerian model has four free parameters: $P$, $T_0$, $\gamma$, and $K_{2}$. We find $P = 0.2279817(7)$ d, $T_0 = 57604.01487(66)$ d, $\gamma = 6(3)$ km s$^{-1}$, $K_{2} = 324(5)$ km s$^{-1}$. The orbital period is consistent with, but more precise than,
the value found by Salvetti et al.~(2015). This orbital fit has an rms of 26 km s$^{-1}$ and a $\chi^2$/d.o.f. of 61/40 (Figure 6), suggesting the possibility that the model does not fully explain the data. 

In black widow systems, an offset between the center of mass and center of light has been observed due to heating of the secondary by high-energy emission associated with the primary (e.g., van Kerkwijk et al.~2011). For redback systems, this effect should be much smaller, but one might still observe a difference between the $K_2$ for the entire data set and the fit for the night ($\phi = 0.05$ to 0.45) side of the secondary, which is less affected by heating than the day side. A fit to these 18 night side velocities gives $K_2 = 327$ km s$^{-1}$, in excellent agreement with the fit to all phases. Hence, at least on the basis of this test, there is no clear evidence that heating affects the measured velocities.

The other piece of evidence for only mild heating is the spectra themselves, which show only modest spectral variations as a function of phase. Generally the spectra are consistent with a mid-G type classification, perhaps as warm as G3 when the ``day" side of the star is visible ($\phi=0.75$) and as cool as G9 when the ``night" side is observed ($\phi=0.25$). These variations only appear to be $\sim \pm200$ K from the values at quadrature.

A separate possibility to explain the sub-optimal circular fit is an eccentric orbit, which, unlike for the other redbacks, cannot yet be constrained by timing a detected radio pulsar. We repeated the radial velocity fits, dropping the assumption of a circular orbit. We found an eccentricity $e=0.046(15)$ but essentially identical values for $P$, $K_2$, and $\gamma$. This fit has an improved $\chi^2$/d.o.f. of 49/38 and an rms of 23 km s$^{-1}$.

It is worth emphasizing that substantial eccentricity is not expected for redbacks with short periods. In particular, for 3FGL J2039.6--5618, the tidal circularization timescale is only $\sim 10^3$ yr (Zahn 1977), though some evidence for eccentricity has been found for candidate redbacks with short orbital periods (Strader et al.~2014). Given the short circularization timescale for 3FGL J2039.6--5618, it seems likely that the apparent eccentricity is false, and is due to an error in a subset of the radial velocities or to non-standard heating that is present outside of the expected phases. A more exotic possibility is that the system is or was a triple, as has been proposed for the origin of unusual eccentric pulsar binaries such as PSR J1903+0327 (Champion et al.~2008). In any case, the orbit could be much better constrained if the likely neutron star primary were detected as a pulsar in the future.

Assuming the circular fit, we find a mass function $f(M) = 0.80(4) M_{\odot}$. If we assume the primary is a neutron star, then $M_{1}$ is likely in the range 1.4 to 2.0 $M_{\odot}$. The measured mass function and the assumed range of $M_{1}$ implies $i \gtrsim 48^{\circ}$, but better constraints come from the light curve fitting below.

\subsubsection{Optical Light Curve Modeling}

Salvetti et al.~(2015) obtained multi-filter photometry for 3FGL J2039.6--5618 in $griz$ using the GROND instrument on the 2.2-m MPG/ESO telescope, and used these data to show the properties were consistent with the secondary of a compact binary. Here we use these same photometric data, combined with our new measurements of the binary properties, in new light curve modeling to obtain improved constraints on the inclination and component masses. 

As discussed by Salvetti et al. (2015), the optical light curves of 3FGL J2039.6--5618 are clearly double-peaked, suggesting that ellipsoidal variations dominate over possible heating from the pulsar. However, the amplitudes of the two peaks at quadrature are asymmetric, and the light curve minimum at inferior conjunction of the primary ($\phi = 0.75$) is brighter than at superior conjunction, indicating that a more complex model is necessary. Here we show that the light curves can be well-reproduced by adding starspots to an underlying model of a tidally deformed secondary (Figure 7).

In detail, we modeled the system using {\tt ELC} (Orosz \& Hauschildt 2000) to fit the GROND light curves in all four bandpasses, following similar procedures as detailed in Swihart et al.~(2017). 
The fits used the {\tt hammerELC} optimizer, a Markov Chain Monte Carlo (MCMC) method based on {\tt emcee} (Foreman-Mackey et al.~2013). We assumed as constraints the values of $P$ and $K_2$ determined from the optical spectroscopy. We also assumed a circular orbit, as well as a mean intensity-weighted secondary surface $T_{\rm eff} = 5600$ K, as suggested by the spectroscopy (the results are insensitive to modest changes in these values). We fit for free parameters $i$ and $q$, the secondary Roche lobe filling factor $f_2$, a phase shift $\Delta \phi$, and a central irradiating luminosity. We also allowed for the presence of circular starspots, each with a temperature profile that changes linearly towards the spot edges, and varying a location, size, and mean $T_{\rm eff}$. We assume foreground reddening from Schlafly \& Finkbeiner (2011) of $E(B-V) = 0.05$.

We find that the light curves are best fit with two large starspots, separated by about 120$^{\circ}$ of longitude.  Our best-fitting model appears to be an excellent representation of the data, but has a formal $\chi^2$/d.o.f. = 425/196. Hence we also ran fits with the photometric uncertainties scaled by a constant factor such that $\chi^2$/d.o.f. = 1. In both cases the median posterior parameter values are essentially identical, but in the latter case the uncertainties are larger. To be conservative, we cite the results from the latter fits.

\begin{figure}[t]
\includegraphics[width=3.4in]{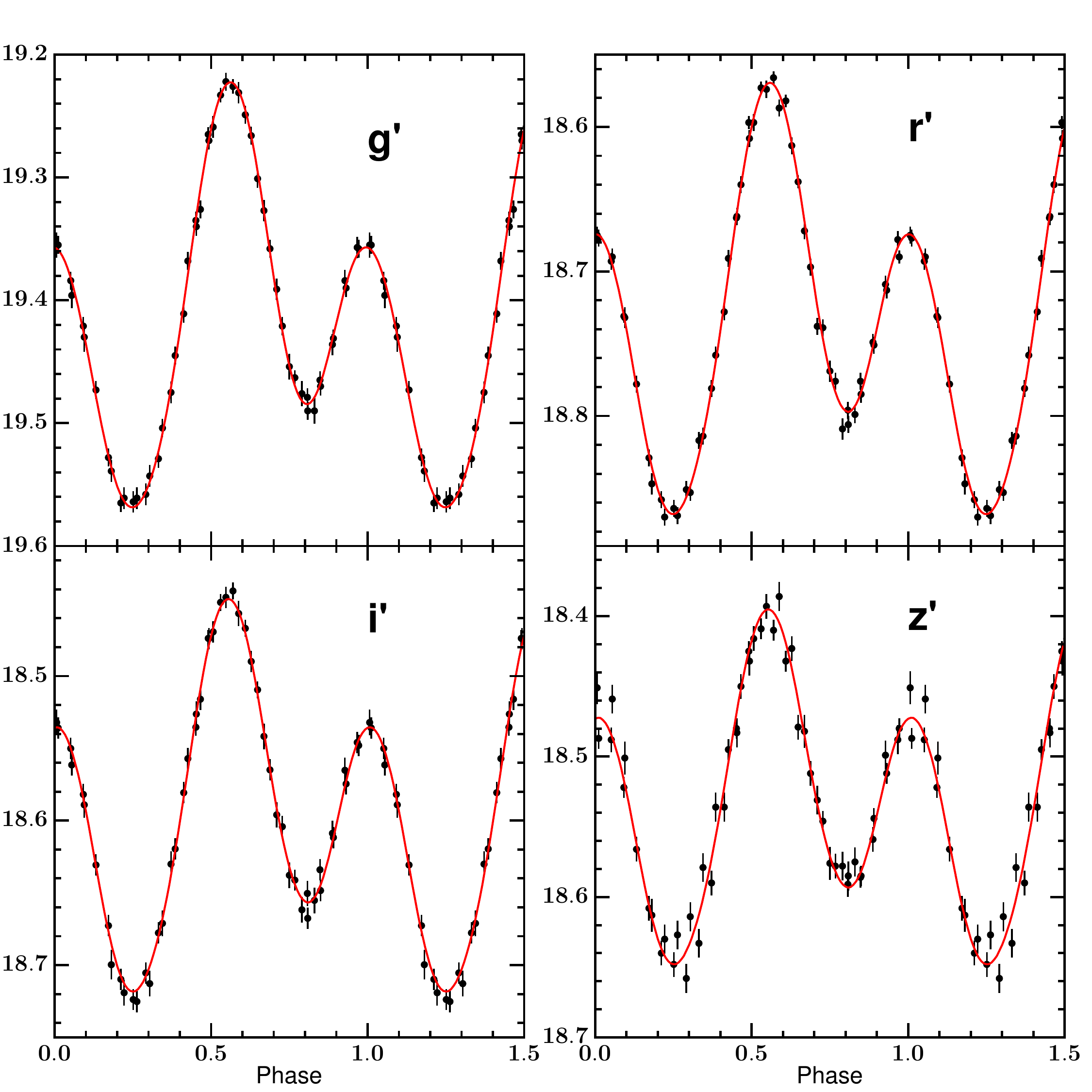}
\caption{{\tt ELC }model fits to the $griz$ photometry for 3FGL J2039.6--5618 (Salvetti et al.~2015), as described in Section 3.5.1.}
\label{fig:j2039_lc}
\end{figure}

We find a best-fitting $i = 57.4^{+2.2^{\circ}}_{-2.3}$ and $q = 0.23^{+0.06}_{-0.04}$. The resulting component masses are $M_{1} = 2.04^{+0.37}_{-0.25} \, M_{\odot}$ and $M_2 = 0.47^{+0.23}_{-0.12} \, M_{\odot}$. The $M_{1}$ value suggests a massive neutron star, consistent with the interpretation of the binary as a redback. We also find $f_2 = 0.95^{+0.04}_{-0.01}$, suggesting that the secondary is close to filling its Roche lobe. We also find a small but significant $\Delta \phi = 0.033(2)$, which corresponds to about a 10.8 min offset between the light curve and the radial velocity curve. Such offsets have been observed in other redbacks (e.g., Li et al~2014), and have been interpreted alternatively as being due to starspots, or due to an offset intrabinary shock. In this best-fit model there is no significant irradiation.

The starspots are both cool ($T_{spot1} = 4700\pm140$; $T_{spot2} = 4120^{+990}_{-340}$ K) and of approximately equal size ($r_{spot1} = 29\pm3^{\circ}$; $r_{spot2} = 29\pm4^{\circ}$). In azimuth they are located at $\theta_{spot1} = 264\pm5^{\circ}$ and $\theta_{spot2} = 148\pm15^{\circ}$, where a value of $180^{\circ}$ corresponds to the night side of the star.

As in Strader et al.~(2016) and Swihart et al.~(2017), we can also estimate the distance to the binary through the normalization of the model light curves compared to the observed source fluxes. We find a distance of $3.4(4)$ kpc. {The individual photometric filters are consistent with a single distance at the $< 1\%$ level, suggesting a physically self-consistent fit to the entire data set.}

Model fits with a single starspot or no starspots, for any value of the irradiation or mass ratio, all give much worse fits than the above two starspot fit. The best alternative model is one with a single starspot and somewhat higher irradiation. This model has $\chi^2$/d.o.f. = 455/199 (original uncertainties) or 210/199 (scaled uncertainties), substantially larger than the two starspot fit with negligible irradiation. Hence, we take the two starspot model as the best current model of the system.

\subsection{Optical Spectra and Emission Lines}

Optical emission lines can provide evidence for an accretion disk, stellar winds, or material ablated from the companion by the pulsar wind. For PSR J1431--4715, PSR J1622--0315, and 3FGL J2039.6--5618, we found no evidence for emission lines at any phase or epoch of our data.

For PSR J1628--3205, H$\alpha$ is detected in absorption at some epochs (2016 Aug 20) and in emission at other epochs (2015 May 16; 2016 Aug 2), but in general it is not apparent, perhaps because it is partially filled in. There is no clear relationship between the presence of H$\alpha$ emission and the orbital phase. The emission, when present, appears to have a velocity consistent with that of the absorption lines. The emission also does not appear to be measurably broadened or double-peaked. The H$\alpha$ emission is likely to be associated with the companion, either directly from its chromosphere, formed in a wind, or perhaps in an intrabinary shock close to the secondary.

However, the SOAR optical spectra for PSR J1048+2339 show perhaps the most extreme emission variations yet observed for any redback. On two different observing nights, 2018 Mar 25 (coverage from $\phi \sim 0.3$--0.5) and 2018 Apr 15 ($\phi \sim 0.05$--0.25), we observe strong, broad emission lines. In some cases the emission lines appear double-peaked and in other epochs only a single component is clearly present. Figure 8 shows an example of one of these spectra, along with spectra of the other sources in our study for reference. In this standard phase convention, at $\phi = 0.25$, the companion is in front of the neutron star along our line of sight. In the epochs where it is most prominent, the extent of the emission ranges from spectra showing only H$\alpha$ to spectra with emission lines from the full Balmer series and \ion{He}{1} at 5876 \AA. The FWHM of the H$\alpha$ emission ranges from $\sim 1200$ to 2400 km s$^{-1}$, and changes substantially on timescales of minutes. By contrast, in data taken on 2018 Feb 22 ($\phi = 0.67$--0.87), when the companion was behind the neutron star, no Balmer emission is apparent in any line (no absorption is observed either, due perhaps to the modest S/N of the spectra or because of infill). Unfortunately, our relatively small collection of observations makes us unable to distinguish between orbital and synoptic variations in the emission. 

We associate the broad and sometimes double-peaked emission lines with material driven off the companion by the pulsar wind, with perhaps an additional contribution from a shock between the companion and the pulsar. Our reasoning is that the emission features change in both strength and width on short timescales ($\lesssim 20$ min), and are not associated with a warm blue continuum as would be expected for an accretion disk. Similar (albeit less extreme) emission features have been observed in the redback/huntsman systems PSR J1740-5340A (in the globular cluster NGC 6397; Sabbi et al.~2003), 3FGL J0838.8--2829 (Halpern et al.~2017b), 1FGL J1417.7--4407 (Strader et al.~2015; Swihart et al.~2018) and in the black widow PSR J1311--3430 (Romani et al.~2015a). 

Such emission features are not observed in all redbacks, and the circumstances in which they appear deserve further study. PSR J1048+2339, which shows the most extreme emission variations, is a typical redback in its mass ratio, orbital period, and even the spindown luminosity of the pulsar (Table 7). It may be noteworthy that Cho et al.~(2018) use optical light curves to show that  the companion of PSR J1048+2339  has undergone repeated episodes of strong, variable heating. It would be beneficial to coordinate optical photometry and spectroscopy to study the connection between the appearance of heating in optical photometry and the emission features in PSR J1048+2339 and other redbacks.

\begin{figure}[t]
\hspace*{-0.7cm}                                                           
\includegraphics[width=3.9in]{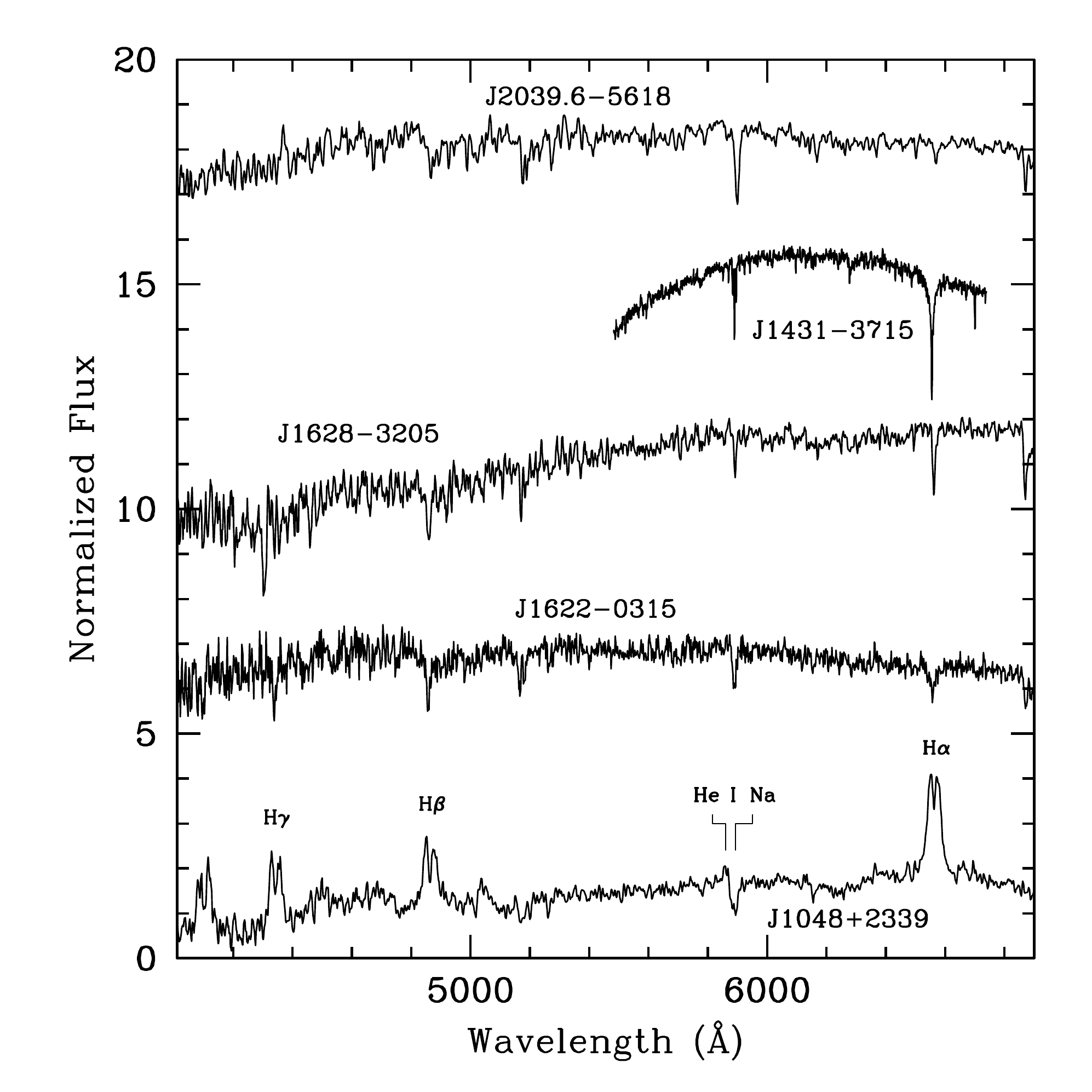}
\caption{Example spectra of the five redbacks with optical spectroscopy in this paper. The spectrum of the companion to PSR J1048+2339 shows the extreme broad emission lines of H and He (the latter blended with Galactic sodium absorption) sometimes observed in this system. Except for PSR J1431--4715, all the spectra are flux-normalized. {The phases of the listed spectra, from bottom to top, are J1048+2339 ($\phi=0.05$), J1622--0315 ($\phi=0.22$), J1628--3205($\phi=0.27$), J1431--4715 ($\phi=0.91$), and J2039.6--5618 ($\phi=0.59$).}
\vspace{1mm}
}
\label{fig:j1048_spec}
\end{figure}

\section{Demographics of Redback Millisecond Pulsars}

\subsection{Defining the Redback Sample}

To properly contextualize the properties of the redbacks that are the subject of this paper, we compiled the available properties of known and candidate redbacks in the Galactic field in Table 7. We divide the sample as follows. Confirmed redbacks are those with a detected millisecond pulsar and a hydrogen-rich companion with a minimum mass of at least $0.1 M_{\odot}$. There are fourteen such systems.
Of these, 1FGL J1417.7--4407 (associated with PSR J1417--4402) has an evolved red giant companion (Strader et al.~2015; Camilo et al.~2016), and we have suggested that this binary be classified as a ``huntsman" rather than a redback, owing to its much longer orbital period, larger secondary, and differing evolutionary fate: unlike typical redbacks, 1FGL J1417.7--4407 is likely to evolve into a typical millisecond pulsar--He white dwarf binary. 

We also include in the ``confirmed" category PSR J1306--40, identified as a candidate redback by Keane et al.~(2018) and Linares (2018) on the basis of the minimum companion mass and a variable optical and X-ray counterpart. A forthcoming paper presents spectroscopic confirmation that this system is a redback, with a hydrogen-rich, {subgiant-like companion} with a minimum mass of $0.49(8) M_{\odot}$ (Swihart et al., in prep.). Pending refined ephemerides from pulsar timing and a more detailed study of its evolution, we leave it in the redback basket at present, but PSR J1306--40 may well be better classified as a huntsman-type system in the future.

To these 13 confirmed redbacks and 1 huntsman-type system, we add 10 candidate redbacks. Two of these contain confirmed pulsars: PSR J1908+2105 and PSR J1302--3258. For PSR J1908+2105, its minimum companion mass of $0.06 M_{\odot}$ would normally lead to a black widow classification, but Cromartie et al.~(2016) argue that most black widows have lower masses, and that the extensive radio eclipses observed in this system are more characteristic of redbacks. Optical photometry and spectroscopy of the faint secondary would allow improved constraints on the secondary mass and an appropriate classification of the binary. PSR J1302--3258 has a minimum companion mass of $0.15 M_{\odot}$ (Hessels et al.~2011), but no optical companion has yet been identified, and no evidence of extensive radio eclipses has yet been published.

The remaining 8 candidates were all discovered through optical and/or X-ray follow-up of \emph{Fermi}-LAT error ellipses of unassociated $\gamma$-ray sources. In all but one case there is an optical source matching the position of an X-ray source in or near the $\gamma$-ray error ellipses, and all 8 systems have follow-up optical photometry and spectroscopy. In 6 of these systems, the properties of the optical source strongly suggest it is the secondary in a binary with a radio millisecond pulsar. Five would be classified as redbacks (3FGL J0212.1+5320, Li et al~2016, Linares et al.~2017;  1FGL J0523.5--2529, Strader et al.~2014; 3FGL J0838.8--2829, Halpern et al.~2017b; 1FGL J0955.2--3949, Li et al.~2018; 3FGL J2039.6--5618, Salvetti et al.~2015 and this paper). The other is a candidate huntsman with an 8.13 d orbital period (2FGL J0846.0+2820; Swihart et al.~2017). 

The final two sources (3FGL J1544.6--1125 and 3FGL J0427.9--6704) show compelling evidence for being low-mass X-ray binaries with ongoing accretion onto a recycled neutron star (Bogdanov \& Halpern 2015; Strader et al.~2016), and X-ray and $\gamma$-ray properties that resemble, at least in some respects, the transitional millisecond pulsars PSR J1023+0038 and XSS J12270--4859 in their sub-luminous disk states (Stappers et al.~2014; de Martino et al.~2013).

There are other systems that have been proposed as possible redbacks, but for which the weight of the evidence currently suggests they are better classified as black widow systems, such as  2FGL J1653.6--0159 (Romani et al.~2014; Kong et al.~2014). These are not included at present, but could be reclassified with future data. There are many remaining unassociated \emph{Fermi}-LAT sources whose properties are consistent with being millisecond pulsars (e.g., Saz Parkinson et al.~2016; Salvetti et al.~2017; Frail et al.~2018); undoubtedly, with follow-up observations, some of these will turn out to be redbacks.

\subsection{A Compilation of Redback Properties}

\subsubsection{Measured Properties}

For all but one system (PSR J1302--3258), previous work on the position of a radio pulsar or optical/X-ray source associated with each system allows the clear identification of a \emph{Gaia} DR2 source (Gaia Collaboration et al.~2018; Gaia Collaboration et al.~2016). We adopt the \emph{Gaia} positions for all these sources. In Table 7 we also list the measured parallax values ($\varpi$), proper motions ($\mu_{\alpha}$ cos $\delta$; $\mu_{\delta}$), and mean $G$ magnitudes, all taken from \emph{Gaia} DR2. Given the preliminary state of \emph{Gaia} astrometry, we have given all these positions a minimum uncertainty of 1 mas per coordinate. Since no precise position for PSR J1302--3258 has yet been published, the position we give is of the brightest \emph{Chandra} X-ray source from v.~2.0 of the \emph{Chandra} Source Catalog (Evans et al.~2010) in the 8-year \emph{Fermi}-LAT error ellipse (radius $\sim 1.5$\arcmin), 2CXO J130225.5--325837. As such, this position should be taken as preliminary. 

The detected millisecond pulsars have typical pulsar parameters listed in Table 7, including the spin period, $\dot{P}$ where available, projected semi-major axis in light-seconds, and dispersion measure.

Compiling distances is important but challenging. The only truly accurate and precise distance is the radio parallax distance for PSR J1023+0038 (Deller et al.~2012). Jennings et al.~(2018) use a pulsar-based distance prior, in combination with \emph{Gaia} DR2 parallax measurements, to determine distances for seven of the confirmed redbacks. Since Jennings et al.~(2018) did not include candidate redbacks in their paper, for those with significant (uncertainty $< 40$\%) parallax measurements, we calculate distances using a simple scale length prior of 1.35 kpc, as suggested by Astraatmadja \& Bailer-Jones (2016). We list all these parallax distances in Table 7 (except for PSR J1023+0038), and generally find that they are consistent with, but often less precise than, the distances from optical light curve fitting. For now we use the optical light curve/spectroscopy distances in most cases, {but recognize the need to revisit and compare optical light curve and dispersion measure-based distances after future \emph{Gaia} data releases.}

There are three exceptions to these guidelines, all among the brightest sources in the sample. 3FGL J0212.1+5320 and PSR J1723--2837 both have precise ($\sim 5$\%) parallax distances from \emph{Gaia} that we use. 1FGL J0523.5--2529 has a $\sim 10$\% precision  \emph{Gaia}  parallax distance ($2.20^{+0.28}_{-0.22}$ kpc) that is larger than the optical distance of $1.1(3)$ kpc (Strader et al.~2014). Unlike most of
the optical light curve distances, which marginalize over the neutron star mass, the 1FGL J0523.5--2529 optical distance assumed a neutron star mass of $1.4 M_{\odot}$. However, if the neutron star is more massive, then the optical distance could be consistent with the parallax distance, which we use. We also note a possible discrepancy for the huntsman candidate 2FGL J0846.0+2820, where the optical distance is large ($8.1\pm1.1$ kpc), while  the \emph{Gaia} distance ($4.4^{+1.3}_{-0.8}$ kpc) is substantially shrunk by the distance prior. We stick with the optical distance pending a future \emph{Gaia} release with a more precise parallax measurement. 

Jennings et al.~(2018) find that the dispersion measure distances from Yao et al.~(2017) appear to be more accurate than those from the Cordes \& Lazio (2002) model for pulsars outside the Plane. Hence, for pulsars with no distance from optical light curve fitting, we adopt the dispersion measure distances that use the Yao et al.~(2017) model, with an assumed 25\% uncertainty.

We take the binary orbital period from whatever method gives the highest precision, which is typically from pulsar timing when available and from optical spectroscopy or photometry otherwise.

Of the 24 known systems, 21 have measurements of the radial velocity semi-amplitude of the secondary ($K_2$) and the systemic velocity $\gamma$. The other three systems have either very faint counterparts (PSR J1908+2105 and PSR J1957+2516) or no optical counterpart yet identified (PSR J1302--3258). Spectroscopy of the counterpart to PSR J1957+2516 will be challenging as a much brighter star is just over 1\arcsec\ away from the source; PSR J1908+2105 should be observable with a 8--10-m class telescope.

Orbital eccentricities are not listed in the catalog as they are small or unmeasured for nearly all systems; the candidate redback 1FGL J0523.5--2529 (Strader et al.~2014) and the candidate huntsman 2FGL J0846.0+2820 (Swihart et al.~2017) have inferred eccentricities of $e=0.040(6)$ and $e=0.061(17)$, respectively. Millisecond pulsar binaries with substantial eccentricities are noteworthy, so it would be desirable to confirm these measurements though the detection of pulsars in the future.

The mass ratios $q = M_c/M_{NS}$ are derived directly from the pulsar and secondary velocity measurements when possible, or from the projected rotational velocity of the optical spectra if no pulsar has yet been detected. 

The binary inclinations are taken from optical light curve fitting if available; else we list constraints assuming a neutron star mass range of 1.4--2.0 $M_{\odot}$. {We note that the inclinations from light curve fitting typically only include statistical uncertainties; the systematic uncertainties, which are generally not quantified, can also be important.} 

When possible, we also list the Roche lobe filling factors inferred from optical light curves; many studies assume that the secondary fills its Roche lobe without explicitly fitting for this parameter. {Only around half of the non-accreting redbacks have estimated filling factors, and many of these are merely notional.}

Table 7 also includes the 0.1--100 GeV $\gamma$-ray and unabsorbed 0.5--10 keV X-ray fluxes (see also Lee et al.~2018). The X-ray fluxes have been homogenized to this common energy range using webPIMMS\footnote{https://heasarc.gsfc.nasa.gov/cgi-bin/Tools/w3pimms/w3pimms.pl} by assuming a power-law model and the best-fit photon index $\Gamma$ in the cited reference, or by using a combined thermal and power-law model if the X-ray study stated such a model was preferred. Finally, we also list the discovery method and the paper that first presented compelling evidence that the source was a redback. We also take the primary ID from these papers, which in most (but not all) cases is a PSR ID for confirmed redbacks and a LAT catalog ID for candidate redbacks. 

\subsubsection{Derived Properties}

We take neutron star and companion mass constraints from the respective papers, or derive such if new information is available in our compilation.
For some objects only lower limits can be listed.

To calculate the intrinsic spindown luminosities of the redbacks detected as pulsars, we use the standard formula: $\dot{E} = 4\pi^2 I \dot{P}/P^3$, where $I = (M_{\rm NS}/1.4 M_{\odot})\, 10^{45}$ g cm$^2$. We correct the observed $\dot{P}$ for the Shklovskii effect: $\dot{P_{sh}}/P = (1/c) (v_t^2/d)$, where $\dot{P_{sh}}$ is the positive contribution from the Shklovskii effect to the observed $\dot{P}$, $v_t$ is the tangential velocity of the system, $c$ is the speed of light, and $d$ is the distance. We calculate $v_t$ using the  \emph{Gaia} proper motions of the binaries and our adopted distance. We use the measured neutron star masses where available or $1.78 M_{\odot}$ otherwise (see \S 4.3.1). This gives typical spindown luminosities about 25\% higher than assuming a standard $1.4  M_{\odot}$ neutron star. The uncertainties in these $\dot{E}$ values includes the uncertainty in $\dot{P}$ and $M_{\rm NS}$, but not the propagated distance uncertainty for the Shklovskii effect.

Using the final adopted distances, we also list the inferred X-ray and $\gamma$-ray luminosities next to the fluxes.

We use the proper motions, distances, and radial velocities to determine the $UVW$ components of the Galactocentric velocity for 21 of the 24 systems.
These individual components are listed in Table 7, assuming the left-handed ($U$ positive toward the Galactic anticenter) convention. 

\subsubsection{Conflicting Measurements}

In most cases, the most accurate published values are apparent, but in other cases conflicting measurements in the literature are worthy of additional discussion.

For PSR J1023+0038, Thorstensen \& Armstrong (2005) find $K_2 = 268(4)$ km s$^{-1}$. The value from McConnell et al.~(2015) is somewhat higher at $K_2 = 286(3)$ km s$^{-1}$, and they argue the higher resolution and better phase sampling of their spectra favor this higher value. For the Roche lobe-filling light curve models of Thorstensen \& Armstrong (2005), which give an inclination of $i = 44(2)^{\circ}$, the implied component masses are $M_{NS} = 1.82(17) M_{\odot}$ and $M_c = 0.24(2) M_{\odot}$. This neutron star mass is slightly higher than found in Deller et al.~(2012), owing entirely to the updated kinematics from McConnell et al.~(2015). However, McConnell et al.~also present  new light curve modeling and suggest that the secondary is underfilling its Roche lobe, resulting in a higher inclination of $i = 54^{+1^{\circ}}_{-5^{\circ}}$ and lower component masses of $M_{NS} = 1.16^{+0.27}_{-0.04} M_{\odot}$  and $M_c = 0.15^{+0.04}_{-0.01} M_{\odot}$. This re-emphasizes the difficulty in accurately determining inclinations for heated low-mass secondaries and the strong dependence of the inferred masses on these inclinations. For this paper we adopt $i=46(2)$, which is within the uncertainties of these measurements, and includes all neutron star masses from 1.4 to 2.0 $M_{\odot}$ within the $2\sigma$ confidence interval.

Regarding XSS J12270--4859, de Martino et al.~(2015) find the photometric constraints on its inclination to yield the range $i = 45$--64$^{\circ}$ (see also Rivera Sandoval et al.~2018). This upper limit would yield, at least for an MSP, an implausibly low neutron star mass of $1.0 M_{\odot}$. Hence, for consistency we list the likely upper inclination in Table 7 as $55^{\circ}$.

PSR J2215+5135 is another system where the inclination is an essential component of the conclusions to the masses of the components. Linares et al.~(2018) find both that (a) the previous radial velocities (Romani et al.~2015b) were affected by heating and thus underestimate the true center of mass motion, and (b) the inclination is more face-on than previous estimates. In combination these results produce a best-fit neutron star mass of $2.27^{+0.17}_{-0.15} M_{\odot}$. By contrast, if the Linares et al.~(2018) radial velocities are used in the context of the intrabinary shock model of Sanchez \& Romani (2017), the inclination is higher and the resulting neutron star mass is $1.93(7) M_{\odot}$, within the range of well-determined masses. These model fits should be revisited in the context of the new observational data available. For now, we list both values in Table 7.

{We emphasize that the uncertainty in Roche lobe filling factors has implications that extend beyond these cited systems. Smaller secondaries show more muted ellipsoidal variations and hence, for a given light curve, imply more edge-on inclinations and lower neutron star masses. Together with the modeling of heating, the filling factor is likely the largest source of uncertainty in determining inclinations from optical light curves. These filling factors also have implications for the optical light curve distances, since for a given light curve smaller stars must be closer. This latter point has the encouraging implication that precise \emph{Gaia} parallax distances in future data releases can independently constrain the companion luminosities and hence sizes.}


\subsection{Masses}

\subsubsection{Neutron Star Masses}

There are 10 systems with direct measurements of the neutron star mass via a combination of optical spectroscopy, optical photometry, and in some cases the projected motion of the pulsar.  An additional four systems have a meaningful lower limit to the neutron mass solely based on the pulsar timing and optical spectroscopy, without any additional constraints on the inclination.

We modeled the neutron star mass distribution of the sample using a hierarchical Bayesian MCMC model. We assumed the neutron stars were drawn from a single normal distribution with a fixed mean (= median) and intrinsic $\sigma$; the model considered the uncertainties in the masses. 

Considering the ten systems with actual mass measurements (not limits), we find a median mass of $1.78\pm0.09 M_{\odot}$, with $\sigma = 0.21\pm0.09$. If we add in the four systems with only lower limits on the primary mass, and use a reasonable upper limit of 2.3 $M_{\odot}$ and a uniform inclination distribution, these values do not meaningfully change. The same goes for the inclusion of PSR J2215+5215, whose principal effect on the analysis is to slightly increase the value of $\sigma$.

Antoniadis et al.~(2016) used precise millisecond pulsar mass estimates to argue for a bimodal mass distribution, with a narrow peak at the canonical mass of $\sim 1.4 M_{\odot}$ and a broad secondary peak at a mass of $1.81^{+0.08}_{-0.13} M_{\odot}$. The redback mass distribution is consistent with being drawn almost entirely from their proposed second peak of massive neutron stars.

There is strong ongoing interest in whether any neutron stars have masses larger than those of PSR J0348+0423 ($2.01(4) M_{\odot}$; Antoniadis et al.~2013) or PSR J1614--2230 ($1.93(2) M_{\odot}$; Fonseca et al.~2016, Demorest et al.~2010). The redback with the highest estimated neutron star mass in our compilation is PSR J2215+5215, with $2.27^{+0.17}_{-0.15} M_{\odot}$ (Linares et al.~2018). As discussed above, this mass is dependent on the model of the heating of the low-mass companion. Of the sources with new data in this paper, the ones most worthy of further attention are the redback PSR J1048+2339, which has a minimum mass of $\geq 1.96(22) M_{\odot}$ independent of light curve modeling ({though the radial velocity precision could be improved}) and the candidate redback 3FGL J2039.6--5618 ($2.04^{+0.37}_{-0.25} M_{\odot}$). There is also an extensive literature on pulsar mass measurements among black widows, of which PSR B1957+20 is perhaps the best-studied, with an inferred mass of $2.40(12) M_{\odot}$ (van Kerkwijk et al.~2011).

Since all the masses significantly above $2 M_{\odot}$ rely on modeling the complex heating of low-mass stars, it cannot yet be claimed that there is compelling evidence for neutron stars with masses significantly above $2 M_{\odot}$. A fair statement is that current observations are consistent with, but do not demand, the existence of such massive neutron stars. 

{Redback neutron star mass measurements typically depend on binary inclinations derived from optical light curve fitting. Some of these systems, especially those with substantial heating of the companion, are affected by additional systematic uncertainties associated with this light curve fitting. This challenge is reflected in the discussion of conflicting measurements above. Despite this caveat, it is worth emphasizing that heating is usually much less important for modeling redbacks than for the less massive black widows. In addition, edge-on systems can offer useful mass constraints independent of of light curve modeling.}

The redback component masses, both neutron star and secondary,  are plotted in Figure 9.

\begin{figure}[t]
\hspace{-0.5cm}
\includegraphics[width=3.8in]{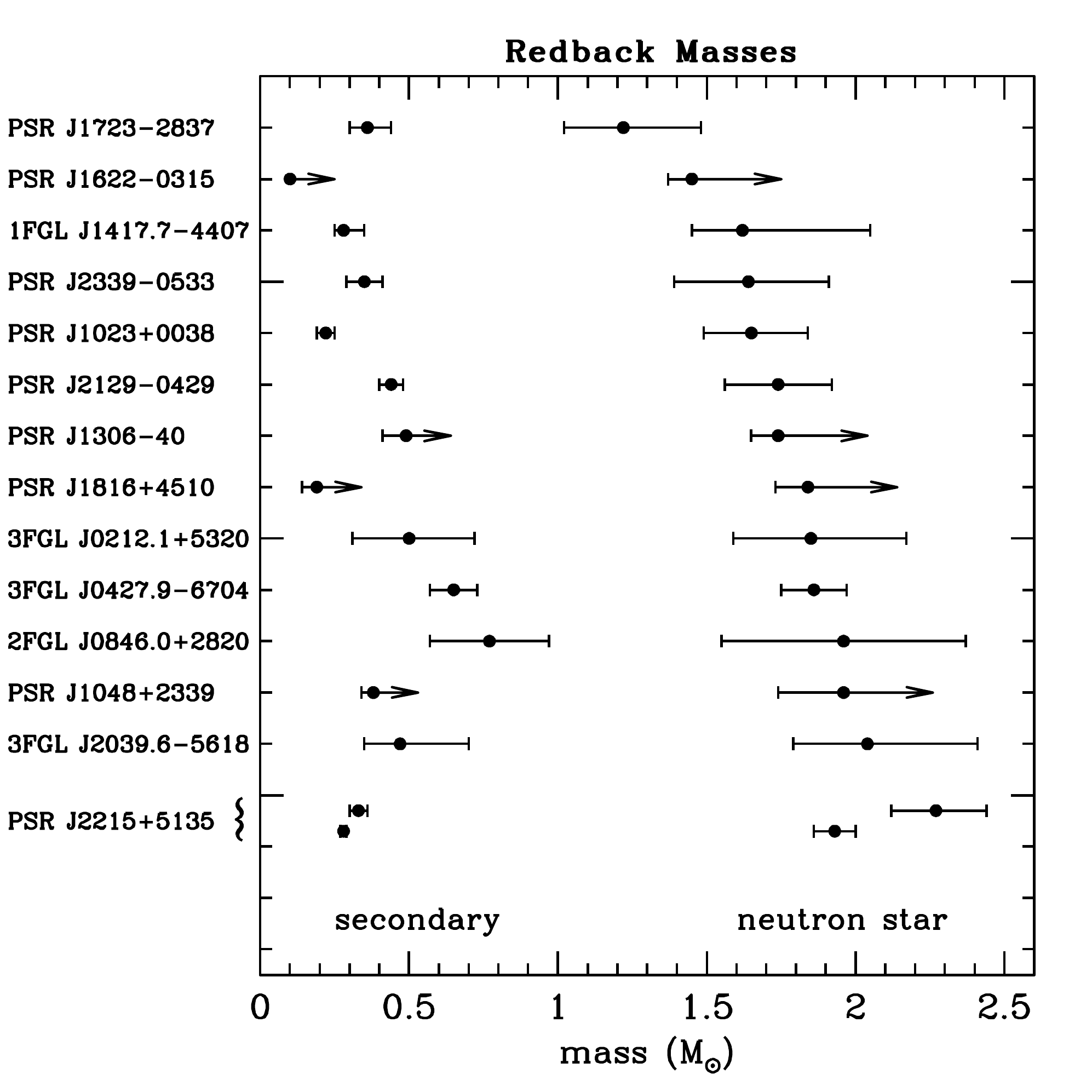}
\caption{Redback component masses for the systems with mass estimates (or lower limits) for the neutron star, sorted by neutron star mass. See similar figure for other neutron star binaries in 	
Lattimer (2012).}
\label{fig:mmm}
\end{figure}

\subsubsection{Companion Star Masses}

Modeling the measured companion masses in the same manner, we find a median mass of $M_c = 0.39\pm0.05 M_{\odot}$, with $\sigma = 0.12\pm0.05$. If we also include the companion mass lower limits (and the same neutron star upper mass limit of 2.3 $M_{\odot}$ as above), these values are $M_c = 0.36\pm0.05 M_{\odot}$ ($\sigma = 0.16\pm0.43$).

These measurements help to refine the parameters of the redback classification. At the low end, PSR J1622--0315 has a companion in the likely mass range 0.10--$0.14 M_{\odot}$, and if PSR J1908+2105 (discussed above) is indeed a redback then it may have a comparably low mass. At the other end of the companion mass scale, among the confirmed redbacks, PSR J1306-40 has a minimum companion mass of $\geq 0.49(8) M_{\odot}$, and PSR J2129--0429 has a companion with a similar mass of $0.44(4) M_{\odot}$. The candidate redbacks 3FGL J0212.1+5320 and 3FGL J2039.6-5618 are both close to $0.5 M_{\odot}$ (with large uncertainties), and the sub-luminous accreting system 3FGL J0427.9--6704 has a precise companion mass of $0.65(8) M_{\odot}$, well-measured because of the eclipses. 1FGL J0523.5--2529 has a high $q$ inferred from optical spectroscopy, which implies a minimum companion mass of $\geq 0.85(8) M_{\odot}$; this unusual system also has a measured eccentricity of $e=0.040(6)$ (Strader et al.~2014). The detection of a pulsar in this system would allow a more direct measurement of these extreme inferred parameters and should be a high priority. In any case, the companion masses in redbacks certainly reach 0.5--$0.6 M_{\odot}$ and likely even larger values.

\subsubsection{Mass Ratios}

The mass ratios are determined independently, and typically with better uncertainties, than the individual neutron star or companion star masses. 17 of the systems have measurements of the mass ratio $q$. For the the 12 redbacks with direct mass ratio measurements from the projected motion of the neutron star and its companion, the median mass ratio is $q=0.16$, with a range of $0.07$ to 0.29. It is notable that among the candidate redbacks in which no pulsar has yet been detected, the mass ratios (inferred from high-resolution optical spectroscopy to determine the projected rotational velocity) are all larger than typical for confirmed redbacks: $q\gtrsim 0.26(3)$ (3FGL J0212.1+5320; Linares et al.~2017); $q=0.61(6)$ (1FGL J0523.5--2529; Strader et al.~2014); and $q=0.40(4)$ for the huntsman candidate 2FGL J0846.0+2820 (Swihart et al.~2017). The mass ratio inferred for the eclipsing low-mass X-ray binary 3FGL J0427.9--6704 is also relatively large at $q=0.35(3)$. In this system the precision is high and systematics minimized owing to the direct detection of both emission and absorption lines from the accretion disk itself (Strader et al.~2016).

There is no evidence that these high mass ratios among candidate redbacks are due to systematic errors in the optical-only mass ratios: for example, Strader et al.~(2015) found  $q =0.18(1)$ for 1FGL J1417.7--4407, while the value from $K_2$ and radio timing is more precise but entirely consistent: $q=0.171(2)$ (Camilo et al.~2016). Instead, it is possible that a bias is present, such that the redbacks with more massive companions (and hence with larger mass ratios) are even harder to detect than typical redbacks. Since the angular size of a near-Roche lobe filling companion as seen from the pulsar increases monotonically with $q$, systems with larger mass ratios could be expected to intercept a higher fraction of the high-energy emission from the pulsar and/or shock, leading to more extensive eclipses. In reality, the situation is likely to be more complicated. High-energy emission from an intrabinary shock could originate closer to the secondary, and its location will vary depending on the properties of the pulsar wind and the magnetic field and stellar wind of the secondary (e.g., Roberts et al.~2014; Romani \& Sanchez 2016). Additional radio timing searches of the candidate redbacks are highly worthwhile.

\begin{figure*}[t]
\hspace*{-0.8cm}                                                           
\includegraphics[width=7.5in]{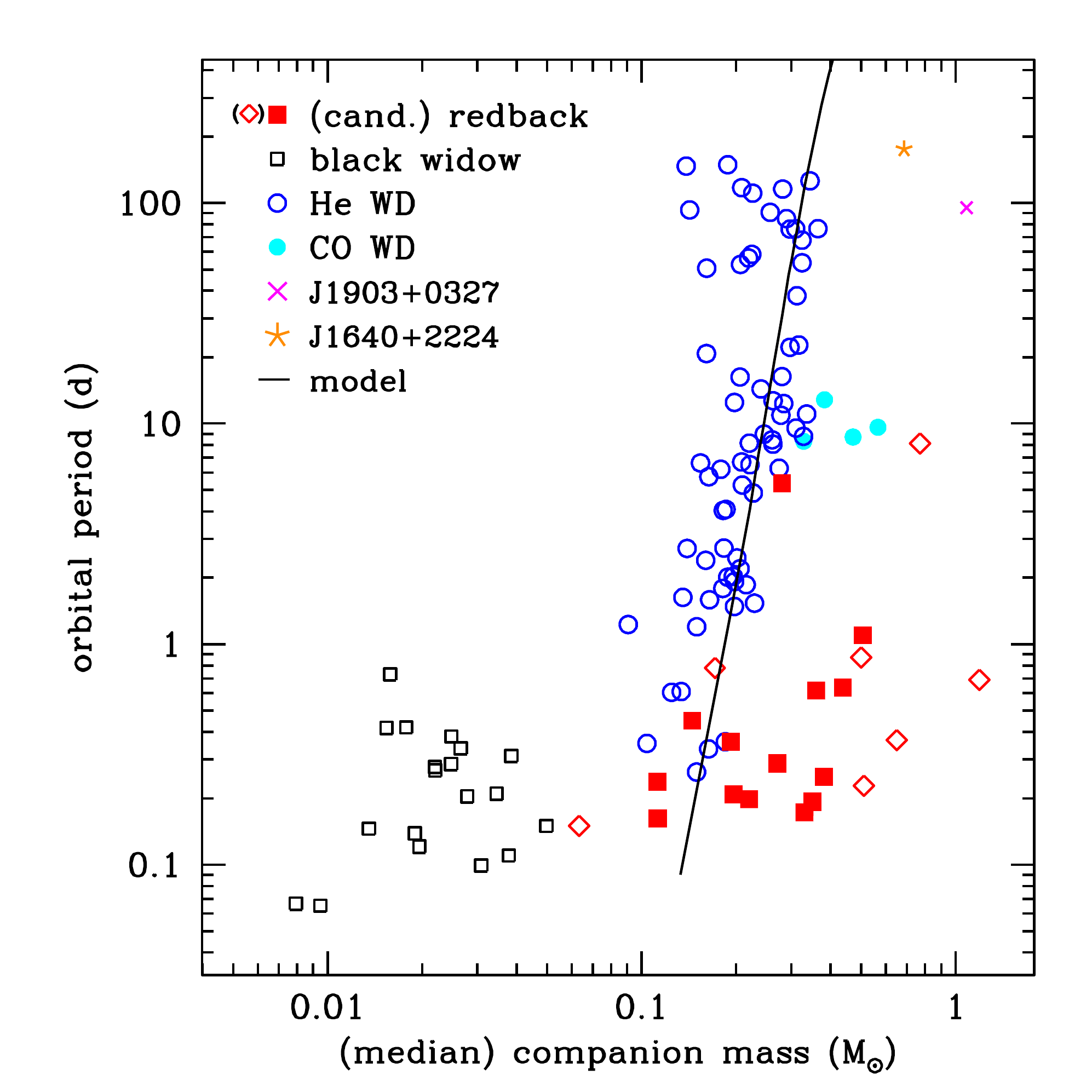}
\caption{\normalsize Orbital period vs.~companion mass for recycled field millisecond pulsars with known companion star types. Most millisecond pulsars have He white dwarfs (open blue circles) from binary evolution, well-represented by models from Tauris \& Savonije (1999). These models (black line) assume solar metallicity and an initial secondary mass of $1.0 M_{\odot}$, and denote the endpoints of an ensemble of systems with varying initial period, not the evolution of a single binary. The CO white dwarfs (filled cyan circles) likely began as close binaries that experienced common-envelope evolution. The field redbacks (filled red squares), candidate redbacks (open red diamonds), and black widows (open black squares) are visible at short orbital period, and it is clear that most of these systems will not have normal field millisecond pulsars as their progeny. The exceptions are the ``huntsman"-type systems 1FGL J1417.7--4407 (Strader et al.~2015; Camilo et al.~2016) and  2FGL J0846.0+2820 (Swihart et al.~2017), which have red giant secondaries. Except for the redbacks with more precise measurements or constraints, the companion masses are ``median" masses that assumed a $1.35 M_{\odot}$ neutron star and an inclination of 60$^{\circ}$, and only recycled systems with spin period $< 8$ msec are plotted. The unusual pulsar binaries PSR J1903+0327 (Champion et al.~2008) and PSR J1640+2224 (Vigeland et al.~2018) are plotted as the magenta cross and orange star, respectively. Figure inspired by Roberts (2013), primarily using data from 
the ATNF Pulsar Catalogue v.1.59 (Manchester et al.~2005).}
\label{fig:rv}
\end{figure*}

\subsection{Kinematics}

Unlike for young pulsars, whose velocities can reflect natal kicks, the bulk velocities of the redbacks should primarily reflect the evolution of these systems in the Galactic potential over long timescales rather than the birth velocity (e.g., Gonzalez et al.~2011; Matthews et al.~2016).

The median three-dimensional Galactocentric velocity is 127 km s$^{-1}$ (with a corresponding one-dimensional velocity of 73 km s$^{-1}$). These values are generally consistent with those of typical millisecond pulsar binaries (e.g., Gonzalez et al.~2011). Using the scaled median absolute deviation to reduce sensitivity to outliers, we find an equivalent $\sigma_{U} = 85$ km s$^{-1}$, $\sigma_{V} = 51$ km s$^{-1}$, and $\sigma_{W} = 43$ km s$^{-1}$. The most extreme individual velocity is that of the candidate transitional millsecond pulsar 3FGL J1544.6--1125, with a three-dimensional velocity of 453 km s$^{-1}$. 

\subsection{Orbital Periods and Evolution}

Figure 10 shows the orbital period vs.~companion mass for millisecond pulsars with known companions (see also Roberts 2013). 

It is well-established that most field millisecond pulsars are fully recycled, such that accretion has permanently ended, and have low-mass He white dwarf companions (Tauris \& van den Heuvel 2006). Figure 10 shows that the relationship between orbital period and white dwarf mass for these systems are well-described by standard binary evolution models (Tauris \& Savonije 1999). Black widows and redbacks deviate from these model predictions, likely due to feedback from the neutron star during the accretion process and/or from a pulsar wind/shock once accretion has ceased (e.g., Chen et al.~2013; Benvenuto et al.~2014). The existence of the transitional millisecond pulsar subclass of redbacks shows that this feedback process can be cyclical. 

Chen et al.~(2013) argue that the distinction between redbacks and black widows is the efficiency of the irradiation, possibly due to beaming, and argue that the evolution of redbacks ``stalls" at masses $> 0.1 M_{\odot}$, such that they do not evolve into black widows. By contrast, Benvenuto et al.~(2014) suggest that short period redbacks (those with orbital periods $< 0.25$ d) will indeed evolve into black widows. Present observations do not clearly distinguish between these possibilities. However, we note that even with the ongoing discovery of new systems, the companion mass distribution of black widows and redbacks is still strongly bimodal (Figure 10). Hence, a model that posits that redbacks typically evolve into black widows must explain, either through speed of evolution or selection effects, why companion masses of $\sim 0.05$--$0.1 M_{\odot}$ seem to be so rare.

It is worth emphasizing that redbacks themselves are not especially rare. They appear to be approximately as common as black widow systems, and depending on whether candidate redbacks are included, make up $\sim 12$--21\% of fully recycled binary millisecond pulsars for which the identity of the companion is known in the ATNF pulsar database (Manchester et al.~2005).
 
Figure 10 also shows the large gap between normal redbacks and the huntsman-type systems 1FGL J1417.7--4407 (Strader et al.~2015; Camilo et al.~2016) and  2FGL J0846.0+2820 (Swihart et al.~2017). 1FGL J1417.7--4407 appears to be in the late stages of the recycling process that will lead to a normal He white dwarf--millisecond pulsar binary, while (if it does indeed contain a pulsar) 2FGL J0846.0+2820 is rather earlier on in the process.

From Figure 10 it is also apparent that there is no difference in the distribution of orbital periods between the confirmed redbacks and candidate redbacks. As discussed above, the candidate systems may have more massive companions, perhaps due to a selection effect on the detection of a radio pulsar in redbacks with more massive companions.

\subsection{High-energy Emission}

In Figure 11 we plot the pulsar spindown power $\dot{E}$ against the 0.1--100 GeV $\gamma$-ray luminosity for the 10 redbacks with appropriate data to calculate these quantities. As discussed above, in all cases the $\dot{E}$ value is corrected for the Shklovskii effect and uses the actual mass of the neutron star rather than assuming $1.4 M_{\odot}$. The median efficiency $\eta = L_{\gamma} / \dot{E} $ is $\sim 10\%$, and considering the uncertainties, most of the redbacks plotted are consistent with this efficiency with a modest dispersion.

Consistent with previous work on larger samples of millisecond pulsars (e.g., Abdo et al.~2013), there is no clear correlation between X-ray and $\gamma$-ray flux for redbacks. In addition, there is no  difference between the X-ray and $\gamma$-ray luminosity distributions of confirmed and candidate redbacks.

\begin{figure}[t]
\includegraphics[width=3.6in]{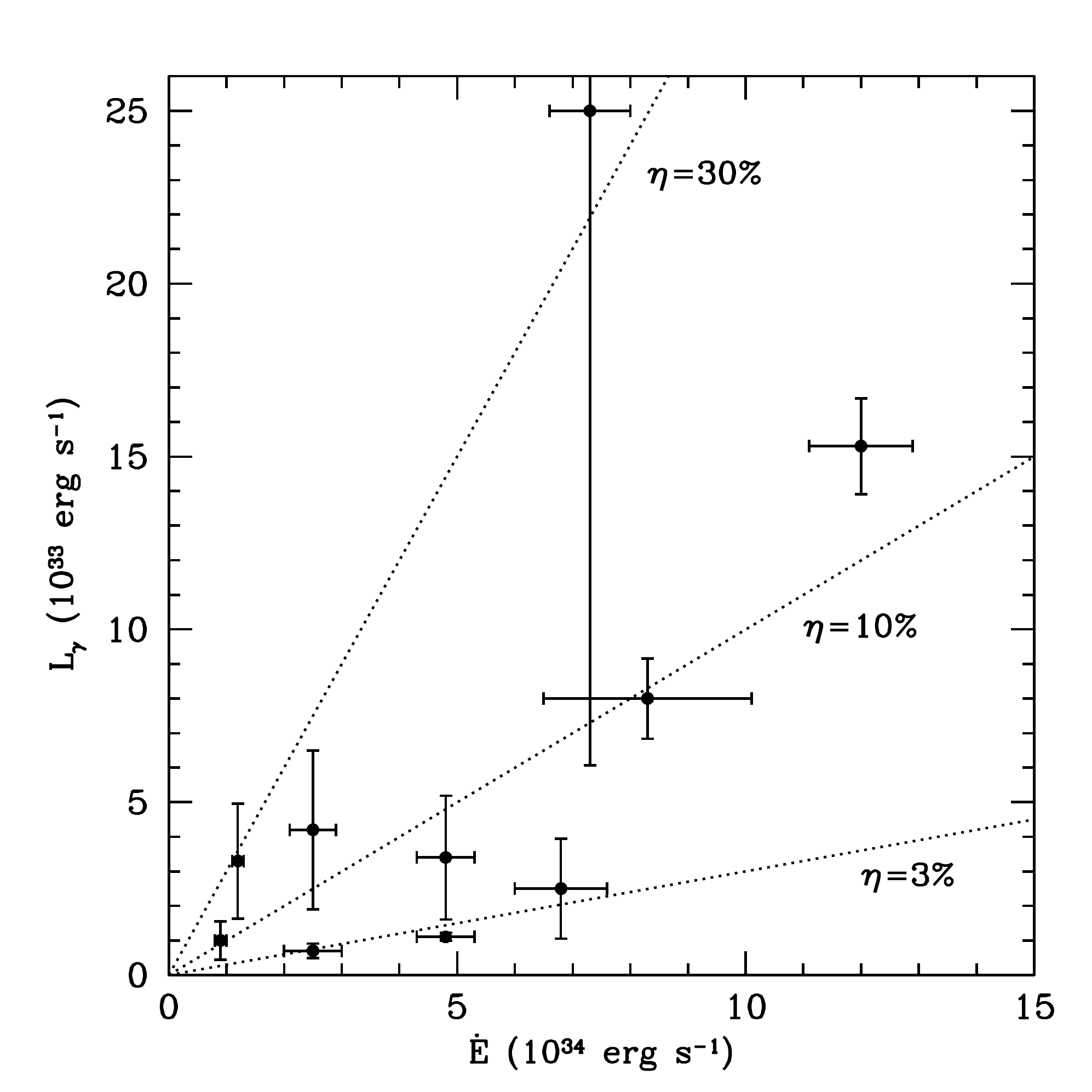}
\caption{$\gamma$-ray luminosity vs spindown power for redback pulsars. The median efficiency $\eta$ of spindown energy conversion to $\gamma$-rays is 10\% with some scatter around this value. The high point with the large uncertainty in its $L_{\gamma}$ is PSR J1816+4510. PSR J1023+0038 and XSS J12270--4859 are plotted in the pulsar state; in the disk state their $L_{\gamma}$ is higher by factors of $\sim 6$ and 2.3, respectively. Unlike the entries in Table 7, the plotted  $L_{\gamma}$ values here include the uncertainties in the distance.}
\label{fig:edot}
\end{figure}

\subsection{Conclusions and Future Work}

Optical spectroscopy has now been obtained for the companions of nearly all confirmed or candidate redback millisecond pulsars. We have used these data to show that the neutron stars in redbacks are typically more massive than the canonical value of $1.4 M_{\odot}$, with a median mass of $1.78\pm0.09 M_{\odot}$ with a dispersion of $\sigma = 0.21\pm0.09$. Several redbacks contain neutron stars whose masses may well be in excess of $2 M_{\odot}$. Most companion stars have masses in the typically quoted range of 0.1--0.5 $M_\odot$, but there is mounting evidence for a subset of redbacks with more massive secondaries, in the range $M_c = 0.5$--0.9 $M_{\odot}$. 
 
It is worth considering how complete current surveys for redbacks are. The median distance of confirmed redbacks is 1.8 kpc, and when excluding the huntsman systems only a few redbacks or candidates have distances $> 3$ kpc. In addition, new candidates with distances of 1--2 kpc have been discovered in just the last 2 years. Hence it is safe to conclude that the redback census is highly incomplete beyond 2 kpc, and X-ray, optical, and radio follow-up of existing and new \emph{Fermi}-LAT sources (from the forthcoming 4FGL catalog) will continue to yield a substantial return. The huntsman systems, with their evolved counterparts, have been discovered at larger distances and may indeed be truly uncommon.

Forthcoming \emph{Gaia} data releases should provide accurate parallax distances for many redbacks. This will enable the systematic improvement of optical light curve measurements of orbital inclinations, resulting in more precise neutron star masses.

Important open questions about redbacks remain, such as the relationship between redbacks and black widows, and whether all redbacks show transitional behavior. The abundance of redbacks and the wealth of information now available on their masses, orbital periods, kinematics, and other properties suggests the timing is propitious to tackle these essential questions through renewed observational and theoretical work.

\acknowledgments

We thank an anonymous referee for helpful comments that improved the paper.  We also thank J.~Antoniadis and M.~Linares for useful discussions. We gratefully acknowledge support from National Science Foundation grant AST-1714825, NASA grants NNX15AU83G and 80NSSC17K0507, and the Packard Foundation. ET acknowledges support from the UnivEarthS Labex program of Sorbonne Paris Cit{\'e} (ANR-10-LABX-0023 and ANR-11-IDEX-0005-02).

Based on observations obtained at the Southern Astrophysical Research (SOAR) telescope, which is a joint project of the Minist\'{e}rio da Ci\^{e}ncia, Tecnologia, e Inova\c{c}\~{a}o (MCTI) da Rep\'{u}blica Federativa do Brasil, the U.S. National Optical Astronomy Observatory (NOAO), the University of North Carolina at Chapel Hill (UNC), and Michigan State University (MSU). This work has made use of data from the European Space Agency (ESA) mission {\it Gaia} (\url{https://www.cosmos.esa.int/gaia}), processed by the {\it Gaia} Data Processing and Analysis Consortium (DPAC, \url{https://www.cosmos.esa.int/web/gaia/dpac/consortium}). Funding for the DPAC has been provided by national institutions, in particular the institutions participating in the {\it Gaia} Multilateral Agreement. This research has made use of data obtained from the Chandra Source Catalog, provided by the Chandra X-ray Center (CXC) as part of the Chandra Data Archive.

{}

\begin{deluxetable}{crr}
\tablecaption{Modified Barycentric Radial Velocities of PSR J1048+2339 \label{tab:bigg}}
\tablehead{MBJD & radial vel. & unc. \\
                   (d)  & (km s$^{-1}$) & (km s$^{-1}$) }
\startdata
58171.1942112 & 157.7 & 27.1 \\
58171.2082393 & 49.3 & 24.4 \\
58171.2283788 & --127.0 & 23.6 \\
58171.2424119 & --308.9 & 27.8 \\
58192.1069703 & --227.1 & 22.3 \\
58202.1674589 & 81.1 & 29.1 \\
58202.1814454 & 195.9 & 23.0 \\
58202.2010592 & 379.1 & 28.4 \\
58202.2150454 & 327.9 & 26.4 \\
58223.1468170 & --390.1 & 28.2 \\
58223.1608030 & --299.4 & 28.1 \\
58223.1832736 & --166.4 & 25.7 \\
58223.1972598 & --5.8 & 29.3
\enddata
\end{deluxetable}

\begin{deluxetable}{crr}
\tablecaption{Modified Barycentric Radial Velocities of PSR J1431--4715 \label{tab:bigg}}
\tablehead{MBJD & radial vel. & unc. \\
                   (d)  & (km s$^{-1}$) & (km s$^{-1}$) }

\startdata
58139.2969337 & --359.8 & 7.8 \\
58139.3109293 & --329.9 & 10.4 \\
58140.3135960 & 22.8 & 6.1 \\
58140.3310646 & 79.6 & 7.6 \\
58160.2049978 & 172.4 & 7.5 \\
58160.2225207 & 148.1 & 6.3 \\
58160.2503889 & 81.8 & 6.0 \\
58160.2680087 & 17.5 & 5.5 \\
58161.2264575 & --195.9 & 5.4 \\
58161.2439261 & --253.4 & 7.1 \\
58161.2682202 & --321.5 & 6.4 \\
58161.2856885 & --343.9 & 6.8 \\
58161.3102045 & --374.5 & 6.9 \\
58161.3276729 & --370.9 & 7.7 \\
58171.2622091 & --291.3 & 7.4 \\
58171.2797871 & --224.6 & 6.2
\enddata
\end{deluxetable}

\begin{deluxetable}{crr}
\tablecaption{Modified Barycentric Radial Velocities of PSR J1622--0315 \label{tab:bigg}}
\tablehead{MBJD & radial vel. & unc. \\
                   (d)  & (km s$^{-1}$) & (km s$^{-1}$) }

\startdata
58016.9924453 & 251.6 & 15.1 \\
58017.0030368 & 294.1 & 15.1 \\
58017.0159592 & 249.0 & 17.7 \\
58017.0264990 & 144.5 & 16.7 \\
58017.0410254 & --70.4 & 14.9 \\
58029.9847623 & --215.7 & 16.3 \\
58029.9953176 & --347.6 & 16.3 \\
58161.3472578 & --349.7 & 22.1 \\
58161.3577815 & --224.9 & 24.1 \\
58161.3757822 & 70.8 & 16.6 \\
58171.3677207 & --419.9 & 16.4 \\
58223.2568012 & --510.9 & 21.4 \\
58223.2721407 & --482.7 & 15.7 \\
58223.2826560 & --343.2 & 14.3 
\enddata
\end{deluxetable}

\begin{deluxetable}{crr}
\tablecaption{Modified Barycentric Radial Velocities of PSR J1628--3205 \label{tab:bigg}}
\tablehead{MBJD & radial vel. & unc. \\
                   (d)  & (km s$^{-1}$) & (km s$^{-1}$) }

\startdata
57158.3475263 & --186.4 & 22.9 \\
57158.3547675 & --277.2 & 23.8 \\
57166.2848526 & --349.7 & 21.8 \\
57170.2906609 & 0.4 & 24.3 \\
57186.3120530 & --86.7 & 24.4 \\
57186.3227294 & 27.9 & 32.3 \\
57196.2693415 & --375.8 & 32.0 \\
57196.2800189 & --278 & 21.7 \\
57252.1511534 & 344.5 & 23.0 \\
57252.1618497 & 327.1 & 26.4 \\
57276.1122885 & 218.9 & 34.5 \\
57276.1229808 & 143.4 & 36.0 \\
57602.1640066 & --289.4 & 23.8 \\
57602.1746989 & --183 & 21.8 \\
57620.0956681 & 0.8 & 20.1 \\
57629.0087887 & --304.2 & 25.9 \\
57629.0195021 & --277.9 & 29.4 
\enddata

\end{deluxetable}

\begin{deluxetable}{crr}
\tablecaption{Modified Barycentric Radial Velocities of 3FGL J2039.6--5618 \label{tab:bigg}}
\tablehead{MBJD & radial vel. & unc. \\
                   (d)  & (km s$^{-1}$) & (km s$^{-1}$) }

\startdata
57307.0672416 & 299.9 & 22.1 \\
57508.4070340 & 216.5 & 21.0 \\
57508.4224817 & 114.5 & 21.6 \\
57598.1874568 & 275.1 & 23.3 \\
57598.1946901 & 270.5 & 21.4 \\
57598.2019655 & 324.1 & 22.6 \\
57598.2161441 & 291.6 & 22.2 \\
57598.2268346 & 279.1 & 23.1 \\
57598.2758183 & --122.5 & 21.2 \\
57598.2865089 & --193.3 & 21.5 \\
57598.2994981 & --297.5 & 22.5 \\
57598.3101950 & --309.6 & 20.9 \\
57598.3232356 & --321.1 & 22.1 \\
57598.3339580 & --298.7 & 22.4 \\
57598.3482067 & --239.7 & 21.3 \\
57598.3745793 & 69.6 & 21.7 \\
57598.3852916 & 159.8 & 22.7 \\
57598.4050533 & 298.0 & 21.6 \\
57602.2639945 & 147.6 & 22.1 \\
57602.2747595 & 211.0 & 21.8 \\
57603.2635285 & 107.5 & 21.4 \\
57603.2742537 & 7.2 & 21.2 \\
57603.2889600 & --136.4 & 21.4 \\
57603.3031855 & --248.3 & 20.9 \\
57620.2355689 & --161.6 & 23.4 \\
57620.2497372 & --64.1 & 22.9 \\
57620.2663429 & 98.3 & 25.9 \\
57629.3299833 & --316.9 & 21.8 \\
57629.3441546 & --267.7 & 21.0 \\
57630.2338185 & --303.3 & 21.5 \\
57630.2479966 & --269.1 & 21.7 \\
57630.3205165 & 231.3 & 21.9 \\
57630.3347184 & 289.3 & 22.7 \\
57630.3489310 & 325.5 & 24.6 \\
57642.1915049 & 328.2 & 21.6 \\
57642.2057022 & 324.8 & 21.9 \\
57642.2219902 & 259.1 & 21.9 \\
57642.2361769 & 213.8 & 22.8 \\
58065.0493676 & 35.5 & 24.4 \\
58071.0733469 & 198.7 & 23.2 \\
58071.0873539 & 103.4 & 23.4 \\
58072.0372005 & --319.2 & 22.1 \\
58072.0692706 & --319.7 & 27.4 \\
58072.1014618 & --151.2 & 23.5
\enddata
\end{deluxetable}

\begin{deluxetable}{crrr}
\tablecaption{SMARTS Photometry of PSR J1431--4715 }
\tablehead{MBJD & Band & Vega Mag & unc. \\
                   (d)  &  & (mag) & (mag) }

\startdata
58143.32592 & B & 18.227 & 0.023 \\
58143.33267 & V & 17.800 & 0.019 \\
58143.33713 & I & 17.230 & 0.027 \\
58144.31750 & B & 18.351 & 0.030 \\
58144.32421 & V & 17.920 & 0.026 \\
58144.32867 & I & 17.295 & 0.038 \\
58145.32596 & B & 18.194 & 0.025 \\
58145.33268 & V & 17.778 & 0.025 \\
58145.33713 & I & 17.184 & 0.033 \\
58146.33884 & B & 18.337 & 0.023 \\
... & ... & ... & ...
\enddata
\tablecomments{Table 6 is published in its entirety in the machine-readable format. A portion is shown here for guidance regarding its form and content.
These magnitudes are not corrected for extinction.}
\label{tab:SMARTSdata}
\end{deluxetable}


\clearpage
\begin{landscape}
\begin{deluxetable*}{llllrrrlrllrrr}
\tablecaption{Observed and Derived Properties of Redbacks \label{tab:redb}}
\tablehead{
ID & other ID & R.A.\tablenotemark{a}  (J2000)                                         & Dec. (J2000)                           & $\varpi$  &     $\mu_{\alpha}$ cos $\delta$ &  $\mu_{\delta}$            &       $G$    &   state\tablenotemark{c}      \\
    &                &    (h:m:s)                                                                            &  ($^{\circ}:\arcmin:\arcsec$)    &    (mas)   &    (mas)  & (mas)               &   (mag)      &           }
\startdata 
PSR J1023+0038   & 2FGL J1023.6+0040 & 10:23:47.68403(7)                  & +00:38:41.007(1)  & 0.73(14)     & $4.75\pm0.14$   & $-17.35\pm0.14$  & 16.265(25) &  PSR/disk \\
PSR J1048+2339    & 3FGL J1048.6+2338 & 10:48:43.43490(7)                  & +23:39:53.579(1)  & 0.96(81)     & $-16.28\pm1.01$ & $-11.70\pm1.27$  & 19.653(28) &  PSR    \\ 
XSS J12270--4859  & PSR J1227--4853 & 12:27:58.7476(1)                   & --48:53:42.826(1)  & 0.62(17)     & $-18.73\pm0.21$ & $7.39\pm0.12$    & 18.076(13) &  PSR/disk \\
                                &      3FGL J1227.9--4854 &                                         &                              &                   &                             &                             &                   &                 \\
PSR J1306--40          & FL8Y J1306.8--4035 & 13:06:56.28036(9)       & --40:35:23.461(1)  & 0.11(28)     & $-5.84\pm0.36$  & $3.88\pm0.33$    & 18.125(15) &  PSR      \\
1FGL J1417.7--4407 & PSR J1417--4402 & 14:17:30.57320(9)             & --44:02:57.498(1)  & 0.221(72)    & $-4.70\pm0.10$  & $-5.10\pm0.09$   & 15.787(8)  &  PSR     \\
PSR J1431--4715  & FL8Y J1431.5--4711 & 14:31:44.6317(1)           & --47:15:27.400(1)  & 0.64(17)     & $-12.01\pm0.33$ & $-14.51\pm0.26$  & 17.750(4)  &  PSR      \\
PSR J1622--0315   & 3FGL J1622.9--0312 & 16:22:59.64071(7)        & --03:15:37.313(1)  & 0.41(52)     & $-13.06\pm1.04$ & $2.96\pm0.68$    & 19.272(11) &  PSR     \\
PSR J1628--3205 & 3FGL J1628.0--3203 & 16:28:07.01033(8)        & --32:05:48.704(1)  & 1.20(56)     & $-6.40\pm1.08$  & $-19.81\pm0.82$  & 19.518(10) &  PSR     \\
PSR J1723--2837  & FL8Y J1722.8--2851\tablenotemark{b} & 17:23:23.19372(8)    & --28:37:57.211(1)  & 1.077(55)    & $-11.71\pm0.08$ & $-23.99\pm0.06$  & 15.549(4)  &  PSR    \\  
PSR J1816+4510 & 3FGL J1816.5+4512  & 18:16:35.93446(9)        & +45:10:33.918(1)  & 0.22(15)     & $-0.17\pm0.29$  & $-4.42\pm0.33$   & 18.221(2)  &  PSR     \\ 
PSR J1957+2516 & \nodata & 19:57:34.61703(7)                            & +25:16:02.194(1)  & 0.69(86)     & $-5.68\pm1.13$  & $-8.86\pm1.47$   & 20.296(7)  &  PSR     \\
PSR J2129--0429 & 3FGL J2129.6--0427  & 21:29:45.04663(7)     & --04:29:06.974(1)  & 0.424(88)    & $12.38\pm0.15$  & $10.19\pm0.15$   & 16.838(7)  &  PSR      \\
PSR J2215+5135 & 3FGL J2215.6+5134  & 22:15:32.6864(1)          & +51:35:36.406(1)  & 0.28(37)     & $0.31\pm0.54$   & $1.88\pm0.60$    & 19.241(22) &  PSR     \\
PSR J2339--0533 & 3FGL J2339.6--0533  & 23:39:38.74105(7)        & --05:33:05.108(1)  & 0.75(26)     & $4.15\pm0.48$   & $-10.31\pm0.31$  & 18.970(71) &  PSR     \\
3FGL J0212.1+5320 & \nodata & 02:12:10.4773(1)                          & +53:21:38.777(1)  & 0.837(35)    & $-2.56\pm0.07$  & $2.10\pm0.06$    & 14.308(5)  &  PSR(?)   \\
1FGL J0523.5--2529 & \nodata & 05:23:16.92771(7)                         & --25:27:37.053(1)  & 0.428(50)    & $3.05\pm0.07$   & $-4.61\pm0.08$   & 16.553(5)  &  PSR(?) \\
3FGL J0838.8--2829 & \nodata & 08:38:50.41805(8)                         & --28:27:56.780(1)  & 0.65(55)     & $-0.01\pm0.66$  & $-12.02\pm0.65$  & 20.035(13) &  PSR(?) \\
2FGL J0846.0+2820 & \nodata & 08:46:21.87570(8)                         & +28:08:40.833(1)  & 0.183(53)    & $-0.46\pm0.08$  & $-2.88\pm0.05$   & 15.661(1)  &  PSR(?)  \\
3FGL J0954.8--3948 & \nodata & 09:55:27.82106(9)                         & --39:47:52.395(1)  & 0.33(17)     & $-8.91\pm0.22$  & $6.38\pm0.25$    & 18.539(23) &  PSR(?)  \\
PSR J1302--3258 & 3FGL J1302.3--3259 & 13:02:25.52(8)              & --32:58:37.0(1)   & \nodata      & \nodata         & \nodata          & \nodata    &  PSR    \\
PSR J1908+2105 & P7R4 J1909+2102 & 19:08:57.28946(7)             & +21:05:02.215(1)  & $-1.6\pm1.3$ & $2.36\pm1.92$   & $-6.38\pm3.46$   & 20.854(9)  &  PSR   \\
3FGL J2039.6--5618  & \nodata & 20:39:34.9603(1)                            & --56:17:09.037(1)  & 0.40(23)     & $4.21\pm0.29$   & $-14.93\pm0.26$  & 18.550(4)  &  PSR(?) \\
3FGL J0427.9--6704 & \nodata & 04:27:49.6117(2)                          & --67:04:35.066(1)  & 0.379(73)    & $12.63\pm0.14$  & $0.31\pm0.16$    & 17.702(20) &  disk      \\
3FGL J1544.6--1125 & \nodata & 15:44:39.36699(7)                         & --11:28:04.684(1)  & 0.60(27)     & $20.11\pm0.53$  & $-12.26\pm0.37$  & 18.610(10) &  disk     \\

\enddata
\tablenotetext{a}{The coordinates, parallax, proper motion, and mean $G$ magnitudes are all are from \emph{Gaia} DR2 (Gaia Collaboration et al.~2018), excepting that for PSR J1302--3258 (see \S 4.2.1).}
\tablenotetext{b}{This association is uncertain.}
\tablenotetext{c}{The accretion state of the binary: PSR (rotation-powered) or disk (accretion powered). Both are listed for the transitional systems. The PSR(?) are systems that
appear to be in the rotation-powered state, but for which no pulsar has yet been detected.}
\end{deluxetable*}
\clearpage
\end{landscape}


\clearpage
\begin{landscape}
\hspace*{-2cm}                                                           
\begin{deluxetable*}{llllrrlllll}
\setcounter{table}{7}
\tablehead{
ID & $P_{\rm spin}$       & $\dot{P}_{obs}$ & $a$ sin $i$   & DM                   & ref\tablenotemark{d} &  CL02\tablenotemark{e}  & Y17\tablenotemark{f}  & other dist. & ref & final dist.\tablenotemark{g} \\
     &       (ms)                 &   $10^{-20}$ s     &   (lt-s)          &  cm$^{-3}$ pc    &                                  &   (kpc)                               & (kpc)                         &  (kpc)    &   & (kpc)  }
\startdata 
PSR J1023+0038    & 1.6879874440059(4)  & 0.683(5)  & 0.3433494(3)   & 14.3 & 18;19  &  0.62    & 1.11    & 1.37(4)                     & 19(PSR $\varpi$)           & 1.37(4) \\
PSR J1048+2339    & 4.6651629362643(16) & 3.00(2)   & 0.836122(3)    & 16.7 & 2  &      0.70    & 2.00    & \nodata                        & \nodata              & 2.0(5) \\
XSS J12270--4859   & 1.68637541026(1)    & 1.11(14)  & 0.668468(4)    & 43.4 & 21  &    1.37    & 1.24    & 1.9(1); $1.37^{+0.69}_{-0.15}$ & 39(opt); 53(\emph{Gaia} $\varpi$) & 1.9(1) \\
PSR J1306--40         & 2.20453(2)          & \nodata   & \nodata        & 35.0 & 22  &         1.20    & 1.41    & 4.7(5)                        & 34              & 4.7(5) \\
1FGL J1417.7--4407  & 2.6642160(4)        & \nodata   & 4.876(9)       & 55.0 & 24  &   1.60    & 2.16    & 3.1(6); $3.50^{+2.09}_{-0.38}$ & 35(opt); 53(\emph{Gaia} $\varpi$) & 3.1(6) \\
PSR J1431--4715    & 2.0119534425332(9)  & 1.411(3)  & 0.550061(2)    & 59.4 & 3  &     1.57    & 1.82    & $1.35^{+0.69}_{-0.15} $          & 53(\emph{Gaia} $\varpi$)          & 1.8(5) \\
PSR J1622--0315    & 3.845429067931(3)   & 1.16(1)   & 0.219258(5)    & 21.4 & 4  &     1.11    & 1.14    & \nodata                        & \nodata              & 1.1(3) \\
PSR J1628--3205    & 3.21                & \nodata   & 0.41026894(41) & 42.1 & 5;52;64  &   1.25    & 1.22    & \nodata                        & \nodata              & 1.2(3) \\
PSR J1723--2837   & 1.855732795728(8)   & 0.75(4)   & 1.225807(9)    & 19.9 & 7  &      0.74    & 0.72    & 0.90(5)                        & 53(\emph{Gaia} $\varpi$)          & 0.90(5) \\
PSR J1816+4510  & 3.1931035538505(2)  & 4.310(1)  & 0.595405(1)    & 38.9 & 8  &        2.42    & 4.36    & $4.5\pm1.7$                    & 37(opt)              & $4.5\pm1.7$ \\
PSR J1957+2516   & 3.961655342404(1)   & 2.744(9)  & 0.283349(6)    & 44.1 & 9  &       3.07    & 2.66    & \nodata                        & \nodata              & 2.7(7) \\
PSR J2129--0429   & 7.62                & \nodata   & 1.855(43)      & 16.9 & 10  &     0.91    & 1.39    & 1.83(11); $2.06^{+0.67}_{-0.21}$ & 10(opt); 53(\emph{Gaia} $\varpi$) & 1.83(11) \\
PSR J2215+5135    & 2.609619723446(1)   & 3.34(7)   & 0.468141(13)   & 69.2 & 11;12  &  3.00    & 2.77    & 2.9(1)                       & 43(opt)              & 2.9(1) \\
PSR J2339--0533    & 2.8842267415473(2)  & 1.4102(6) & 0.611656(4)    & 8.7  & 15  &    0.45    & 0.75    & 1.1(3); $1.08^{+0.82}_{-0.12}$ & 38(opt); 53(\emph{Gaia} $\varpi$) & 1.1(3) \\
3FGL J0212.1+5320 & \nodata             & \nodata   & \nodata        & \nodata & \nodata  &     \nodata & \nodata & $0.92^{+0.12}_{-0.16}$;1.16(5)   & 42(opt); 29(\emph{Gaia} $\varpi$) & 1.16(5) \\
1FGL J0523.5--2529  & \nodata             & \nodata   & \nodata        & \nodata & \nodata  &   \nodata & \nodata & 1.1(3); $2.20^{+0.28}_{-0.22}$ & 27(opt); 29(\emph{Gaia} $\varpi$) & 2.2(3) \\
3FGL J0838.8--2829  & \nodata             & \nodata   & \nodata        & \nodata & \nodata  &   \nodata & \nodata & 1.0(?)                         & 28(opt)              & 1.0(3) \\
2FGL J0846.0+2820  & \nodata             & \nodata   & \nodata        & \nodata & \nodata  &    \nodata & \nodata & $8.1\pm1.1$; $4.40^{+1.26}_{-0.84}$ & 30(opt); 29(\emph{Gaia} $\varpi$) & $8.1\pm1.1$ \\
3FGL J0954.8--3948  & \nodata             & \nodata   & \nodata        & \nodata & \nodata  &   \nodata & \nodata & 1.7(7)                         & 31(X-ray/opt)         & 1.7(7) \\
PSR J1302--3258    & 3.77                & \nodata   & 0.928          & 26.2 & 11  &    1.00    & 1.43    & \nodata                        & \nodata              & 1.4(4) \\
PSR J1908+2105    & 2.56                & \nodata   & 0.12           & 61.9 & 16  &     3.18    & 2.60    & \nodata                        & \nodata              & 2.6(6) \\
3FGL J2039.6--5618    & \nodata             & \nodata   & \nodata        & \nodata & \nodata  &         \nodata & \nodata & 3.4(4)                         & 29(opt)              & 3.4(4) \\
3FGL J0427.9--6704   & \nodata             & \nodata   & \nodata        & \nodata & \nodata  &  \nodata & \nodata & 2.4(3); $2.46^{+0.59}_{-0.40}$ & 32(opt); 29(\emph{Gaia} $\varpi$) & 2.4(3) \\
3FGL J1544.6--1125   & \nodata             & \nodata   & \nodata        & \nodata & \nodata  &  \nodata & \nodata & 3.8(7)                       & 33(opt)              & 3.8(7) \\
\enddata
\tablenotetext{d}{Reference for radio pulsar properties.}
\tablenotetext{e}{Distance using the pulsar dispersion measure and the Cordes \& Lazio (2002) electron density model.}
\tablenotetext{f}{Distance using the pulsar dispersion measure and the Yao et al.~(2017) electron density model.}
\tablenotetext{g}{Final adopted distance.}
\end{deluxetable*}
\clearpage
\end{landscape}

\clearpage

\begin{landscape}
\hspace*{-4cm}                                                           
\begin{deluxetable*}{llllrlrrr}
\hspace*{-4cm}      
\setcounter{table}{7}                                                     
\tablecaption{Observed and Derived Properties of Redbacks \label{tab:redb3}}
\tablehead{
 ID &       orb.~P          & ref &   $K_2$\tablenotemark{h}           & $\gamma$\tablenotemark{i}         & ref      & $U$               & $V $              &       $W$        \\
     &           (d)             &      &  (km s$^{-1}$)  & (km s$^{-1}$)   &           & (km s$^{-1}$) & (km s$^{-1}$) & (km s$^{-1}$)   }  
\startdata 
PSR J1023+0038 &   0.1980963569(3) & 20  & 286(3)       & 0(2)     & 41      & --88(1)  & --65(1)    & --27(2)     \\
PSR J1048+2339  & 0.250519045(6)  & 2   & 376(14)      & --24(8)   & 29      & 74(10)  & --125(12)  & --97(8)      \\
XSS J12270--4859    &    0.287887519(1)  & 21  & 261(5)       & 67(2)    & 39      & 118(2)  & --117(2)   & 71(1)       \\
PSR J1306--40 &         1.09716(6)      & 23  & 210(2)       & 32(2)    & 34      & 95(7)   & --55(5)    & 106(7)        \\
1FGL J1417.7--4407 &    5.37372(3)      & 24  & 116(1)       & --15(1)   & 36      & 52(1)   & --54(1)    & --43(1)      \\
PSR J1431--4715   &     0.4497391377(7) & 3   & 278(3)       & --91(2)   & 29      & 139(2)  & --49(2)    & --86(2)        \\
PSR J1622--0315  &      0.1617006798(6) & 4   & 423(8)       & --135(6)  & 29      & 134(6)  & --40(4)    & --6(5)       \\  
PSR J1628--3205  &      0.2081445828    & 6   & 358(10)      & --4(7)    & 29      & 9(7)    & --93(5)    & --43(6)    \\    
PSR J1723--2837  &      0.615436473(8)  & 7   & 148(2)       & 33(2)    & 40      & --38(2)  & --101(1)   & --7(1)   \\      
PSR J1816+4510 &        0.3608934817(2) & 8   & 343(7)       & --99(8)   & 37      & --73(7)  & --92(8)    & --53(7) \\        
PSR J1957+2516 &        0.2381447210(7) & 9   & \nodata      & \nodata  & \nodata & \nodata & \nodata   & \nodata \\
PSR J2129--0429  &      0.6352274131(3) & 10  & 250(4)       & --64(2)   & 10      & 142(2)  & 40(2)     & 8(2)          \\
PSR J2215+5135  &       0.172502105(8)  & 12  & 412(5)       & 49(8)    & 43      & 18(8)   & 60(8)     & 22(8)        \\
PSR J2339--0533  &      0.1930984018(3) & 15  & 320(15)      & --49(8)   & 38      & --12(2)  & --60(4)    & 24(7)        \\
3FGL J0212.1+5320  &    0.869575(4)     & 26  & 214(5)       & -8(5)    & 42      & --25(4)  & 20(4)     & 14(1)        \\
1FGL J0523.5--2529  &   0.688134(28)    & 27  & 190(1)       & 57(1)    & 27      & --12(1)  & --66(1)    & --9(1)        \\
3FGL J0838.8--2829  &   0.214507(5)     & 28  & 315(17)      & 129(15)  & 28      & --8(6)   & --127(14)  & --10(4)      \\ 
2FGL J0846.0+2820  &    8.13284(43)     & 30  & $54.4\pm1.0$ & 43(1)    & 30      & 21(2)   & --102(2)   & --4(3)      \\  
3FGL J0954.8--3948 &    0.3873396(81)   & 31  & 272(4)       & 96(3)    & 31      & 80(2)   & --82(3)    & 20(2)     \\   
PSR J1302--3258 &       0.784           & 17  & \nodata      & \nodata  & \nodata & \nodata & \nodata   &  \nodata \\
PSR J1908+2105  &       0.15            & 16  & \nodata      & \nodata  & \nodata & \nodata & \nodata   & \nodata \\
3FGL J2039.6--5618 &    0.2279817(7)    & 29  & 324(5)       & 6(3)     & 29      & 78(3)   & --187(4)   & --27(4)       \\
3FGL J0427.9--6704 &    0.3667200(7)    & 32  & 293(4)       & 79(3)    & 32      & 27(2)   & --138(3)   & 60(2)         \\
3FGL J1544.6--1125 &    0.2415361(36)   & 33  & $39.3\pm1.5$ & 144(1)   & 33      & --360(5) & 78(8)     & --263(7)       \\
\enddata
\tablenotetext{h}{Semi-amplitude $K_2$ of the companion star.}
\tablenotetext{i}{Systemtic velocity $\gamma$ of the binary.}
\end{deluxetable*}
\clearpage
\end{landscape}

\begin{landscape}
\hspace*{-4cm}                                                           
\begin{deluxetable*}{lllrrllllll}
\hspace*{-4cm}      
\setcounter{table}{7}                                                     
\tablecaption{Observed and Derived Properties of Redbacks \label{tab:redb4}}
\tablehead{
 ID &        $M_{c}/M_{NS}$ & ref & $M_{NS}$             & $M_{c}$          & ref & inc.\tablenotemark{j}     & ref & $f$\tablenotemark{k} & ref  \\
       &                                  &       &  ($M_{\odot}$)      & ($M_{\odot}$) &         & ($^{\circ}$)                  &       &                              &  }  
\startdata 
PSR J1023+0038          & 0.132(1)  & 29       & $1.65^{+0.19}_{-0.16}$ & 0.22(3)      & 29 & 46(2) & 41;59                      & 1;0.83(3) & 59;41 \\
PSR J1048+2339         & 0.193(7)  & 29       & $\geq 1.96$(22)        & $\geq 0.38$(4)    & 29 & $> 83^{+7}_{-10}$ & 29  & \nodata & \nodata   \\
XSS J12270--4859       & 0.194(4)  & 29         & \nodata                & $\geq 0.27$(1)    & 29 & $> 46$(1); $< 55$(2) & 29;61 & $\sim 1$ & 61   \\
PSR J1306--40          & 0.285(30) & 34         & $\geq 1.74$(9)              & $\geq 0.49$(8)    & 29 & $> 73$(4) & 29 & 0.94(5) & 34   \\
1FGL J1417.7--4407     & 0.171(2)  & 29       & $1.62^{+0.43}_{-0.17}$ & $0.28^{+0.07}_{-0.03}$ & 35 & 64(3) & 35 & 0.83(6) & 35 \\
PSR J1431--4715       & 0.096(1)  & 29       & \nodata                & $\geq 0.13$       & 3 & $> 58$(1); $< 72$(2) & 29 & 0.70(6) & 29   \\
PSR J1622--0315        & 0.070(1)  & 29       & $\geq 1.45$(8)              & $\geq 0.10$       & 29 & $> 64$(2) & 29 & \nodata & \nodata   \\
PSR J1628--3205        & 0.120(3)  & 29       & \nodata                & $\geq 0.17$       & 5;52 & $> 59$(3); $< 74$(6) & 29 & \nodata & \nodata   \\
PSR J1723--2837        & 0.293(4)  & 29       & $1.22^{+0.26}_{-0.20}$ & $0.36^{+0.08}_{-0.06}$ & 58;29 & 41(3) & 58 & $\sim 1$ & 58   \\
PSR J1816+4510         & 0.105(2)  & 29      & $\geq 1.84$(11)             & $\geq 0.19$(5)    & 37 & $> 77$(5) & 29 & $< 1$ & 37 \\
PSR J1957+2516        & \nodata   & \nodata     & \nodata                & $\geq 0.10$       & 9 & \nodata & \nodata & \nodata & \nodata   \\
PSR J2129--0429        & 0.255(4)  & 10         & 1.74(18)               & 0.44(4)      & 10 & 81(7) & 10 &  0.95(1) & 10 \\
PSR J2215+5135          & 0.144(2)  & 29          & $2.27^{+0.17}_{-0.15}$ & 0.33(3) & 29;43 & 64(3) & 43  & 0.95(1)  & 43 \\
                                      &                 &               & 1.93(7)                          & 0.28(1) &    62   &   82(1) & 62  & 0.64(1) & 62 \\
PSR J2339--0533         & 0.216(10) & 29       & $1.64^{+0.27}_{-0.25}$ & 0.35(6)      & 15;29 & 57(2) & 38 & 0.90(1) &  38 \\
3FGL J0212.1+5320       & 0.28(8)   & 60         & $1.85^{+0.32}_{-0.26}$ & $0.50^{+0.22}_{-0.19}$ & 60 & 69(4) & 60 & 0.76(3) & 60 \\
1FGL J0523.5--2529     & 0.61(6)   & 27       & \nodata                 & $\geq 0.85$(8)    & 27 & $> 59$(2); $< 75$(6) & 29 &  \nodata & \nodata   \\
3FGL J0838.8--2829      & \nodata   & \nodata  & \nodata                & \nodata & \nodata & $> 47$(3) & 29 &  \nodata & \nodata   \\
2FGL J0846.0+2820       & 0.402(36) & 30        & 1.96(41)               & 0.77(20)     & 30 & $31^{+3}_{-2}$ & 30 & $\sim 1$  & 30 \\
3FGL J0954.8--3948     & \nodata   & \nodata    & \nodata                & \nodata & \nodata & $> 50$ & 29 & \nodata & \nodata   \\
PSR J1302--3258        & \nodata   & \nodata     & \nodata                & $\geq 0.15$ & 11 & \nodata & \nodata & \nodata & \nodata   \\
PSR J1908+2105         & \nodata   & \nodata     & \nodata                & $\geq 0.06$ & 16 & \nodata & \nodata & \nodata & \nodata   \\
3FGL J2039.6--5618     & \nodata   & \nodata   & $2.04^{+0.37}_{-0.25}$ & $0.47^{+0.23}_{-0.12}$ & 29 & 57(2) & 29 & $0.95^{+0.04}_{-0.01}$ & 29 \\
3FGL J0427.9--6704     & 0.35(3)   & 32         & 1.86(11)               & 0.65(8) & 32 & 78(2) & 32 & 1 & 32 \\
3FGL J1544.6--1125     & \nodata   & \nodata   & \nodata                & \nodata & \nodata & 7(1) & 33 & 1 & 33 \\
\enddata
\tablenotetext{j}{Measurement of or constraint on the binary inclination $i$.}
\tablenotetext{k}{Roche lobe filling factor $f$.}
\end{deluxetable*}
\clearpage
\end{landscape}

\thispagestyle{empty}
\begin{landscape}
\hspace*{-4cm}                                                           
\begin{deluxetable*}{llllllllllllll}
\hspace*{-4cm}                                 
\setcounter{table}{7}                          
\tablecaption{Observed and Derived Properties of Redbacks \label{tab:redb5}}
\tablehead{
ID   &        $\dot{P}_{corr}$\tablenotemark{l}         & $\dot{E} $                                   &    $F_{\gamma}$\tablenotemark{m}                                  & $L_{\gamma}$           & ref &  $F_{X}$\tablenotemark{n}                                                  & $L_{X}$                   & ref & discovery & ref \\
       &       $10^{-20}$ s            &   ($10^{34}$ erg s$^{-1}$)           &   ($10^{-12}$ erg s$^{-1}$ cm$^{-2}$)   &  ($10^{33}$ erg s$^{-1}$)   &     &    ($10^{-13}$ erg s$^{-1}$ cm$^{-2}$) &  ($10^{31} $erg s$^{-1}$)  &     &                    &  }
\startdata 
PSR J1023+0038\tablenotemark{o}   &        0.50(1)         & 4.8(5)      &  4.9(5)/$29.3\pm2.8$       & 1.1(1)/6.6(6)     & 46         & 7(2)/105(2)         & 16(4)/236(4)   & 57 & GBT survey & 18 \\
PSR J1048+2339 &        2.09(15)        & 1.2(1)      &  6.9(5)                    & 3.3(2)            & 44         & $0.63(10) $              & $3.0(5)$            & 52  & radio follow-up of \emph{Fermi} & 16;2 \\
XSS J12270--4859\tablenotemark{o}   &      0.79(14)        & $8.3\pm1.8$ &  $18.6\pm1.9$/$41.6\pm1.5$ & 8.0(8)/18.0(7)    & 45         & 6.3(9)/179(4)       & 27(4)/773(17)  & 57 & X-ray survey & 51 \\
PSR J1306--40 &         \nodata         & \nodata     &  13.1(8)                   & 35(2)            & 44         & 5.1(1)              & 135(3)        & 23 & SUPERB/Parkes survey & 22 \\
1FGL J1417.7--4407 &    \nodata         & \nodata     &  12.8(8)                   & 14.7(9)           & 44         & 9.8(5)              & 113(6)         & 35 & opt/X-ray search of \emph{Fermi} & 36 \\
PSR J1431--4715   &     1.10(1)         & 6.8(8)      &  $6.4\pm1.0$               & 2.5(4)            & 44         & \nodata             & \nodata        & \nodata & HTRU/Parkes survey & 3 \\
PSR J1622--0315  &      0.98(4)         & 0.9(1)      &  6.7(8)                    & 1.0(1)            & 44         & 0.22(4)             & 0.32(6)        & 54 & radio follow-up of \emph{Fermi} & 4 \\
PSR J1628--3205  &      \nodata         & \nodata     & $11.1\pm1.1$               & 1.9(2)            & 44         & 1.3(1)              & 2.24(2)        & 56 & radio follow-up of \emph{Fermi} & 5 \\
PSR J1723--2837  &      0.46(4)         & 2.5(5)      &  $7.0\pm1.7$               & 0.7(2)            & 44         & 36(9)               & 35(9)          & 57 & Parkes survey & 7 \\
PSR J1816+4510 &        4.24(1)         & 7.3(7)      &  10.5(6)                   & 25(1)             & 44         & 0.05(1)             & 1.2(2)         & 8 & radio follow-up of \emph{Fermi} & 8 \\
PSR J1957+2516 &        2.46(10)        & 2.0(2)      &  \nodata                   & \nodata           & \nodata &  \nodata               & \nodata        & \nodata & PALFA/Arecibo survey & 9 \\
PSR J2129--0429  &      \nodata         & 4.8(5)\tablenotemark{q}         &  8.4(6)                    & 3.4(3)            & 44         & 2.85(7)             & 11.4(3)        & 56 & radio follow-up of \emph{Fermi} & 10 \\
PSR J2215+5135  &       3.33(7)         & 12.0(9)     &  15.2(9)                   & 15.3(9)           & 44         & 1.00(5)             & 10.1(5)        & 54 & radio follow-up of \emph{Fermi} & 11 \\
PSR J2339--0533  &      1.32(1)         & 2.5(4)      &  $29.3\pm1.0$              & 4.2(2)            & 44         & 1.9(3)              & 2.8(4)         & 57 & opt/X-ray search of \emph{Fermi} & 6;38;50 \\
3FGL J0212.1+5320  &    \nodata         & \nodata     & $15.8\pm1.0$               & 2.5(2)            & 44         & 18(1)               & 29(2)          & 42 & opt/X-ray search of \emph{Fermi} & 26;42 \\
1FGL J0523.5--2529  &   \nodata         & \nodata     &  12.1(7)                   & 7.0(4)            & 44         & 2.4(6)              & 14(3)          & 27 & opt/X-ray search of \emph{Fermi} & 27 \\
3FGL J0838.8--2829  &   \nodata         & \nodata     &  8.6(7)                    & 1.0(1)            & 44         & 1.63(7)             & 1.95(8)        & 49 & opt/X-ray search of \emph{Fermi} & 49 \\
2FGL J0846.0+2820\tablenotemark{p}  &    \nodata         & \nodata     &  $8.5\pm2.7$       & $67\pm21$ & 30 & \nodata             & \nodata        & \nodata & opt/X-ray search of \emph{Fermi} & 30 \\
3FGL J0954.8--3948 &    \nodata         & \nodata     & $10.7\pm1.0$               & 3.7(4)            & 44         & $2.9^{+2.0}_{-0.6}$ & $10.0^{+6.9}_{-2.1}$ & 31 & opt/X-ray search of \emph{Fermi} & 31 \\
PSR J1302--3258 &       \nodata         & \nodata     & 10.5(6)                    & 2.5(2)            & 44         & 0.16(4)             & 0.38(9)        & 25 & radio follow-up of \emph{Fermi} & 11 \\
PSR J1908+2105  &       \nodata         & \nodata     &  $7.0\pm1.1$               & 5.7(9)            & 44         & 0.30(7)             & 2.4(6)         & 54 & radio follow-up of \emph{Fermi} & 16 \\
3FGL J2039.6--5618 &    \nodata         & \nodata     & 15.5(8)                    & 21(1)             & 44         & 0.97(5)             & 13.4(7)        & 48 & opt/X-ray search of \emph{Fermi} & 48 \\
3FGL J0427.9--6704 &    \nodata         & \nodata     &  8.5(6)                    & 5.9(4)            & 44         & 35(2)               & 241(14)        & 32 & opt/X-ray search of \emph{Fermi} & 32 \\
3FGL J1544.6--1125 &    \nodata         & \nodata     & $14.1\pm1.1$               & 24(2)             & 44         & 32.1(4)             & 555(7)         & 55 & opt/X-ray search of \emph{Fermi} & 55 \\
\enddata
\tablenotetext{l}{Spin-down rate of the pulsar, corrected for the Shklovskii effect using the observed proper motion of the companion star and our adopted final distances.}
\tablenotetext{m}{$\gamma$-ray flux from \emph{Fermi}-LAT over the energy range 0.1--100 GeV.}
\tablenotetext{n}{X-ray flux  over the energy range 0.5--10 keV. The listed reference is the source of the flux; the luminosity is calculated using the adopted final distance from this paper.}
\tablenotetext{o}{$\gamma$-ray and X-ray properties in both the PSR and disk states are given.}
\tablenotetext{p}{The listed $\gamma$-ray flux/luminosity are from the period before mid-2009; at later times this source is not detected in \emph{Fermi}.}
\tablenotetext{q}{See Roberts et al.~(2014).}
\tablecomments{{\bf References} \\
1: \emph{Gaia} Collaboration et al.~(2018); 2: Deneva et al.~(2016); 3: Bates et al.~(2015); 4: Sanpa-Arsa (2016); 5: Ray et al.~(2012);
6: Kong et al.~(2012); 7: Crawford et al.~(2013); 8: Stovall et al.~(2014); 9: Stovall et al.~(2016); 10: Bellm et al.~(2016);
11: Hessels et al.~(2011); 12: Abdo et al.~(2013); 13: Manchester et al.~(2005); 14: Roberts et al.~(2014); 15: Pletsch \& Clark (2015);
16: Cromartie et al.~(2016); 17: D.~Lorimer (http://astro.phys.wvu.edu/GalacticMSPs/GalacticMSPs.txt); 18: Archibald et al.~(2009); 19: Deller et al.~(2012); 20: Archibald et al.~(2013);
21: Roy et al.~(2015); 22: Keane et al.~(2018); 23: Linares (2017); 24: Camilo et al.~(2016); 25: Evans et al.~(2010);
26: Li et al.~(2016); 27: Strader et al.~(2014); 28: Halpern et al.~(2017b); 29: this work; 30: Swihart et al.~(2017);
31: Li et al.~(2018); 32: Strader et al.~(2016); 33: Britt et al.~(2017); 34: Swihart et al., in preparation; 35: Swihart et al.~(2018);
36: Strader et al.~(2015); 37: Kaplan et al.~(2013); 38: Romani \& Shaw (2011); 39: de Martino et al.~(2014); 40: Antoniadis et al.~, in prep.;
41: McConnell et al.~(2015); 42: Linares et al.~(2017); 43: Linares et al.~(2018); 44: \emph{Fermi}-LAT FL8Y (https://fermi.gsfc.nasa.gov/ssc/data/access/lat/fl8y/); 45: Johnson et al.~(2015);
46: Stappers et al.~(2014); 47: Bogdanov (2016); 48: Salvetti et al.~(2015); 49: Halpern et al.~(2017a); 50: Ray et al.~(2014);
51: Hill et al.~(2011); 52: Cho et al.~(2018); 53: Jennings et al.~(2018); 54: Gentile (2018); 55: Bogdanov \& Halpern (2015)
56: Roberts et al.~(2015); 57: Linares (2014); 58: van Staden \& Antoniadis (2016); 59: Thorstensen \& Armstrong (2005); 60: Shahbaz et al.~(2017);
61: de Martino et al.~(2015); 62: Sanchez \& Romani (2017); 63: al Noori et al.~(2018); 64: S.~Ransom (2014, private communication)\\}

\end{deluxetable*}
\clearpage
\end{landscape}


\begin{thebibliography}{}

\bibitem[Abdo et al.(2013)]{2013ApJS..208...17A} Abdo, A.~A., Ajello, M., Allafort, A., et al.\ 2013, \apjs, 208, 17 
\bibitem[Acero et al.(2015)]{2015ApJS..218...23A} Acero, F., Ackermann, M., Ajello, M., et al.\ 2015, \apjs, 218, 23 
\bibitem[Al Noori et al.(2018)]{2018ApJ...861...89A} Al Noori, H., Roberts, M.~S.~E., Torres, R.~A., et al.\ 2018, \apj, 861, 89 
\bibitem[Antoniadis et al.(2013)]{2013Sci...340..448A} Antoniadis, J., Freire, P.~C.~C., Wex, N., et al.\ 2013, Science, 340, 448 
\bibitem[Antoniadis et al.(2016)]{2016arXiv160501665A} Antoniadis, J., Tauris, T.~M., Ozel, F., et al.\ 2016, arXiv:1605.01665 
\bibitem[Archibald et al.(2009)]{2009Sci...324.1411A} Archibald, A.~M., Stairs, I.~H., Ransom, S.~M., et al.\ 2009, Science, 324, 1411 
\bibitem[Archibald et al.(2013)]{2013arXiv1311.5161A} Archibald, A.~M., Kaspi, V.~M., Hessels, J.~W.~T., et al.\ 2013, arXiv:1311.5161 
\bibitem[Astraatmadja \& Bailer-Jones(2016)]{2016ApJ...832..137A} Astraatmadja, T.~L., \& Bailer-Jones, C.~A.~L.\ 2016, \apj, 832, 137 
\bibitem[Bassa et al.(2014)]{2014MNRAS.441.1825B} Bassa, C.~G., Patruno, A., Hessels, J.~W.~T., et al.\ 2014, \mnras, 441, 1825 
\bibitem[Bates et al.(2015)]{2015MNRAS.446.4019B} Bates, S.~D., Thornton, D., Bailes, M., et al.\ 2015, \mnras, 446, 4019 
\bibitem[Bellm et al.(2016)]{2016ApJ...816...74B} Bellm, E.~C., Kaplan, D.~L., Breton, R.~P., et al.\ 2016, \apj, 816, 74 
\bibitem[Benvenuto et al.(2014)]{2014ApJ...786L...7B} Benvenuto, O.~G., De Vito, M.~A., \& Horvath, J.~E.\ 2014, \apjl, 786, L7 
\bibitem[Bogdanov(2016)]{2016ApJ...826...28B} Bogdanov, S.\ 2016, \apj, 826, 28 
\bibitem[Bogdanov \& Halpern(2015)]{2015ApJ...803L..27B} Bogdanov, S., \& Halpern, J.~P.\ 2015, \apjl, 803, L27 
\bibitem[Britt et al.(2017)]{2017ApJ...849...21B} Britt, C.~T., Strader, J., Chomiuk, L., et al.\ 2017, \apj, 849, 21 
\bibitem[Camilo et al.(2016)]{2016ApJ...820....6C} Camilo, F., Reynolds, J.~E., Ransom, S.~M., et al.\ 2016, \apj, 820, 6 
\bibitem[Champion et al.(2008)]{2008Sci...320.1309C} Champion, D.~J., Ransom, S.~M., Lazarus, P., et al.\ 2008, Science, 320, 1309 
\bibitem[Chen et al.(2013)]{2013ApJ...775...27C} Chen, H.-L., Chen, X., Tauris, T.~M., \& Han, Z.\ 2013, \apj, 775, 27 
\bibitem[Cho et al.(2018)]{Cho18} Cho, P., Halpern, J., Bogdanov, S. 2018, \apj, 866, 71
\bibitem[Clemens et al.(2004)]{2004SPIE.5492..331C} Clemens, J.~C., Crain, J.~A., \& Anderson, R.\ 2004, \procspie, 5492, 331 
\bibitem[Cordes \& Lazio(2002)]{2002astro.ph..7156C} Cordes, J.~M., \& Lazio, T.~J.~W.\ 2002, arXiv:astro-ph/0207156 
\bibitem[Crawford et al.(2013)]{2013ApJ...776...20C} Crawford, F., Lyne, A.~G., Stairs, I.~H., et al.\ 2013, \apj, 776, 20 
\bibitem[Cromartie et al.(2016)]{2016ApJ...819...34C} Cromartie, H.~T., Camilo, F., Kerr, M., et al.\ 2016, \apj, 819, 34 
\bibitem[de Martino et al.(2013)]{2013A&A...550A..89D} de Martino, D., Belloni, T., Falanga, M., et al.\ 2013, \aap, 550, A89 
\bibitem[de Martino et al.(2014)]{2014MNRAS.444.3004D} de Martino, D., Casares, J., Mason, E., et al.\ 2014, \mnras, 444, 3004 
\bibitem[de Martino et al.(2015)]{2015MNRAS.454.2190D} de Martino, D., Papitto, A., Belloni, T., et al.\ 2015, \mnras, 454, 2190 
\bibitem[Deller et al.(2012)]{2012ApJ...756L..25D} Deller, A.~T., Archibald, A.~M., Brisken, W.~F., et al.\ 2012, \apjl, 756, L25 
\bibitem[Demorest et al.(2010)]{2010Natur.467.1081D} Demorest, P.~B., Pennucci, T., Ransom, S.~M., Roberts, M.~S.~E., \& Hessels, J.~W.~T.\ 2010, \nat, 467, 1081 
\bibitem[Deneva et al.(2016)]{2016ApJ...823..105D} Deneva, J.~S., Ray, P.~S., Camilo, F., et al.\ 2016, \apj, 823, 105 
\bibitem[Eastman et al.(2010)]{2010PASP..122..935E} Eastman, J., Siverd, R., \& Gaudi, B.~S.\ 2010, \pasp, 122, 935 
\bibitem[Evans et al.(2010)]{2010ApJS..189...37E} Evans, I.~N., Primini, F.~A., Glotfelty, K.~J., et al.\ 2010, \apjs, 189, 37 
\bibitem[Fonseca et al.(2016)]{2016ApJ...832..167F} Fonseca, E., Pennucci, T.~T., Ellis, J.~A., et al.\ 2016, \apj, 832, 167 
\bibitem[Foreman-Mackey et al.(2013)]{2013PASP..125..306F} Foreman-Mackey, D., Hogg, D.~W., Lang, D., \& Goodman, J.\ 2013, \pasp, 125, 306 
\bibitem[Frail et al.(2018)]{2018MNRAS.475..942F} Frail, D.~A., Ray, P.~S., Mooley, K.~P., et al.\ 2018, \mnras, 475, 942 
\bibitem[Gaia Collaboration et al.(2016)]{2016A&A...595A...1G} Gaia Collaboration, Prusti, T., de Bruijne, J.~H.~J., et al.\ 2016, \aap, 595, A1 
\bibitem[Gaia Collaboration et al.(2018)]{2018A&A...616A...1G} Gaia Collaboration, Brown, A.~G.~A., Vallenari, A., et al.\ 2018, \aap, 616, A1 
\bibitem[Gentile(2018)]{2018PhDT.........4G} Gentile, P.~A.\ 2018, Ph.D.~Thesis, West Virginia University
\bibitem[Gonzalez et al.(2011)]{2011ApJ...743..102G} Gonzalez, M.~E., Stairs, I.~H., Ferdman, R.~D., et al.\ 2011, \apj, 743, 102 
\bibitem[Halpern et al.(2017)]{2017ApJ...838..124H} Halpern, J.~P., Bogdanov, S., \& Thorstensen, J.~R.\ 2017a, \apj, 838, 124 
\bibitem[Halpern et al.(2017)]{2017ApJ...844..150H} Halpern, J.~P., Strader, J., \& Li, M.\ 2017b, \apj, 844, 150 
\bibitem[Hessels et al.(2011)]{2011AIPC.1357...40H} Hessels, J.~W.~T., Roberts, M.~S.~E., McLaughlin, M.~A., et al.\ 2011, American Institute of Physics Conference Series, 1357, 40 
\bibitem[Hill et al.(2011)]{2011MNRAS.415..235H} Hill, A.~B., Szostek, A., Corbel, S., et al.\ 2011, \mnras, 415, 235 
\bibitem[Iglesias-Marzoa et al.(2015)]{2015PASP..127..567I} Iglesias-Marzoa, R., L{\'o}pez-Morales, M., \& Jes{\'u}s Ar{\'e}valo Morales, M.\ 2015, \pasp, 127, 567 
\bibitem[Johnson et al.(2015)]{2015ApJ...806...91J} Johnson, T.~J., Ray, P.~S., Roy, J., et al.\ 2015, \apj, 806, 91 
\bibitem[Jennings et al.(2018)]{2018arXiv180606076J} Jennings, R.~J., Kaplan, D.~L., Chatterjee, S., Cordes, J.~M., \& Deller, A.~T.\ 2018, \apj, 864, 26 
\bibitem[Kaplan et al.(2013)]{2013ApJ...765..158K} Kaplan, D.~L., Bhalerao, V.~B., van Kerkwijk, M.~H., et al.\ 2013, \apj, 765, 158 
\bibitem[Keane et al.(2018)]{2018MNRAS.473..116K} Keane, E.~F., Barr, E.~D., Jameson, A., et al.\ 2018, \mnras, 473, 116 
\bibitem[Kong et al.(2012)]{2012ApJ...747L...3K} Kong, A.~K.~H., Huang, R.~H.~H., Cheng, K.~S., et al.\ 2012, \apjl, 747, L3 
\bibitem[Kong et al.(2014)]{2014ApJ...794L..22K} Kong, A.~K.~H., Jin, R., Yen, T.-C., et al.\ 2014, \apjl, 794, L22 
\bibitem[Lattimer(2012)]{2012ARNPS..62..485L} Lattimer, J.~M.\ 2012, Annual Review of Nuclear and Particle Science, 62, 485 
\bibitem[Lee et al.(2018)]{2018ApJ...864...23L} Lee, J., Hui, C.~Y., Takata, J., et al.\ 2018, \apj, 864, 23 
\bibitem[Li et al.(2016)]{2016ApJ...833..143L} Li, K.-L., Kong, A.~K.~H., Hou, X., et al.\ 2016, \apj, 833, 143 
\bibitem[Li et al.(2018)]{2018ApJ...863..194L} Li, K.-L., Hou, X., Strader, J., et al.\ 2018, \apj, 863, 194 
\bibitem[Li et al.(2014)]{2014ApJ...795..115L} Li, M., Halpern, J.~P., \& Thorstensen, J.~R.\ 2014, \apj, 795, 115 
\bibitem[Linares(2014)]{2014ApJ...795...72L} Linares, M.\ 2014, \apj, 795, 72 
\bibitem[Linares(2018)]{2018MNRAS.473L..50L} Linares, M.\ 2018, \mnras, 473, L50 
\bibitem[Linares et al.(2017)]{2017MNRAS.465.4602L} Linares, M., Miles-P{\'a}ez, P., Rodr{\'{\i}}guez-Gil, P., et al.\ 2017, \mnras, 465, 4602 
\bibitem[Linares et al.(2018)]{2018ApJ...859...54L} Linares, M., Shahbaz, T., \& Casares, J.\ 2018, \apj, 859, 54 
\bibitem[Manchester et al.(2005)]{2005AJ....129.1993M} Manchester, R.~N., Hobbs, G.~B., Teoh, A., \& Hobbs, M.\ 2005, \aj, 129, 1993 
\bibitem[Matthews et al.(2016)]{2016ApJ...818...92M} Matthews, A.~M., Nice, D.~J., Fonseca, E., et al.\ 2016, \apj, 818, 92 
\bibitem[McConnell et al.(2015)]{2015MNRAS.451.3468M} McConnell, O., Callanan, P.~J., Kennedy, M., Hurley, D., Garnavich, P., \& Menzies, J.\ 2015, \mnras, 451, 3468
\bibitem[Orosz \& Hauschildt(2000)]{2000A&A...364..265O} Orosz, J.~A., \& Hauschildt, P.~H.\ 2000, \aap, 364, 265 
\bibitem[{\"O}zel \& Freire(2016)]{2016ARA&A..54..401O} {\"O}zel, F., \& Freire, P.\ 2016, \araa, 54, 401 
\bibitem[Pletsch \& Clark(2015)]{2015ApJ...807...18P} Pletsch, H.~J., \& Clark, C.~J.\ 2015, \apj, 807, 18 
\bibitem[Ray et al.(2012)]{2012arXiv1205.3089R} Ray, P.~S., Abdo, A.~A., Parent, D., et al.\ 2012, arXiv:1205.3089 
\bibitem[Ray et al.(2014)]{2014AAS...22314007R} Ray, P.~S., Belfiore, A.~M., Saz Parkinson, P., et al.\ 2014, American Astronomical Society Meeting Abstracts \#223, 223, 140.07 
\bibitem[Rivera Sandoval et al.(2018)]{2018MNRAS.476.1086R} Rivera Sandoval, L.~E., Hern{\'a}ndez Santisteban, J.~V., Degenaar, N., et al.\ 2018, \mnras, 476, 1086 
\bibitem[Roberts(2013)]{2013IAUS..291..127R} Roberts, M.~S.~E.\ 2013, Neutron Stars and Pulsars: Challenges and Opportunities after 80 years, 291, 127 (arXiv:1210.6903)
\bibitem[Roberts et al.(2014)]{2014AN....335..313R} Roberts, M.~S.~E., Mclaughlin, M.~A., Gentile, P., et al.\ 2014, Astronomische Nachrichten, 335, 313 
\bibitem[Roberts et al.(2015)]{2015arXiv150207208R} Roberts, M.~S.~E., McLaughlin, M.~A., Gentile, P.~A., et al.\ 2015, arXiv:1502.07208 
\bibitem[Romani(2015)]{2015ApJ...812L..24R} Romani, R.~W.\ 2015, \apjl, 812, L24 
\bibitem[Romani \& Shaw(2011)]{2011ApJ...743L..26R} Romani, R.~W., \& Shaw, M.~S.\ 2011, \apjl, 743, L26 
\bibitem[Romani et al.(2014)]{2014ApJ...793L..20R} Romani, R.~W., Filippenko, A.~V., \& Cenko, S.~B.\ 2014, \apjl, 793, L20 
\bibitem[Romani et al.(2015a)]{2015ApJ...804..115R} Romani, R.~W., Filippenko, A.~V., \& Cenko, S.~B.\ 2015, \apj, 804, 115 
\bibitem[Romani et al.(2015b)]{2015ApJ...809L..10R} Romani, R.~W., Graham, M.~L., Filippenko, A.~V., \& Kerr, M.\ 2015, \apjl, 809, L10 
\bibitem[Romani \& Sanchez(2016)]{2016ApJ...828....7R} Romani, R.~W., \& Sanchez, N.\ 2016, \apj, 828, 7 
\bibitem[Roy et al.(2015)]{2015ApJ...800L..12R} Roy, J., Ray, P.~S., Bhattacharyya, B., et al.\ 2015, \apjl, 800, L12 
\bibitem[Sabbi et al.(2003)]{2003ApJ...589L..41S} Sabbi, E., Gratton, R., Ferraro, F.~R., et al.\ 2003, \apjl, 589, L41 
\bibitem[Salvetti et al.(2015)]{2015ApJ...814...88S} Salvetti, D., Mignani, R.~P., De Luca, A., et al.\ 2015, \apj, 814, 88 
\bibitem[Salvetti et al.(2017)]{2017MNRAS.470..466S} Salvetti, D., Mignani, R.~P., De Luca, A., et al.\ 2017, \mnras, 470, 466 
\bibitem[Sanchez \& Romani(2017)]{2017ApJ...845...42S} Sanchez, N., \& Romani, R.~W.\ 2017, \apj, 845, 42 
\bibitem[Sanpa-Arsa(2016)]{2016PhDT.........4S} Sanpa-Arsa, S. 2016, Ph.D.~Thesis, University of Virginia
\bibitem[Saz Parkinson et al.(2016)]{2016ApJ...820....8S} Saz Parkinson, P.~M., Xu, H., Yu, P.~L.~H., et al.\ 2016, \apj, 820, 8 
\bibitem[Schlafly \& Finkbeiner(2011)]{2011ApJ...737..103S} Schlafly, E.~F., \& Finkbeiner, D.~P.\ 2011, \apj, 737, 103 
\bibitem[Shahbaz et al.(2017)]{2017MNRAS.472.4287S} Shahbaz, T., Linares, M., \& Breton, R.~P.\ 2017, \mnras, 472, 4287 
\bibitem[Stappers et al.(2014)]{2014ApJ...790...39S} Stappers, B.~W., Archibald, A.~M., Hessels, J.~W.~T., et al.\ 2014, \apj, 790, 39 
\bibitem[Stovall et al.(2014)]{2014ApJ...791...67S} Stovall, K., Lynch, R.~S., Ransom, S.~M., et al.\ 2014, \apj, 791, 67 
\bibitem[Stovall et al.(2016)]{2016ApJ...833..192S} Stovall, K., Allen, B., Bogdanov, S., et al.\ 2016, \apj, 833, 192 
\bibitem[Strader et al.(2014)]{2014ApJ...788L..27S} Strader, J., Chomiuk, L., Sonbas, E., et al.\ 2014, \apjl, 788, L27 
\bibitem[Strader et al.(2015)]{2015ApJ...804L..12S} Strader, J., Chomiuk, L., Cheung, C.~C., et al.\ 2015, \apjl, 804, L12 
\bibitem[Strader et al.(2016)]{2016ApJ...831...89S} Strader, J., Li, K.-L., Chomiuk, L., et al.\ 2016, \apj, 831, 89 
\bibitem[Swihart et al.(2017)]{2017ApJ...851...31S} Swihart, S.~J., Strader, J., Johnson, T.~J., et al.\ 2017, \apj, 851, 31 
\bibitem[Swihart et al.(2018)]{2018arXiv180807101S} Swihart, S.~J., Strader, J., Shishkovsky, L., et al.\ 2018, \apj, 866, 83
\bibitem[Tauris \& Savonije(1999)]{1999A&A...350..928T} Tauris, T.~M., \& Savonije, G.~J.\ 1999, \aap, 350, 928 
\bibitem[Tauris \& van den Heuvel(2006)]{2006csxs.book..623T} Tauris, T.~M., \& van den Heuvel, E.~P.~J.\ 2006, Compact stellar X-ray sources, 39, 623 
\bibitem[Thorstensen \& Armstrong(2005)]{2005AJ....130..759T} Thorstensen, J.~R., \& Armstrong, E.\ 2005, \aj, 130, 759 
\bibitem[van Kerkwijk et al.(2011)]{2011ApJ...728...95V} van Kerkwijk, M.~H., Breton, R.~P., \& Kulkarni, S.~R.\ 2011, \apj, 728, 95 
\bibitem[van Staden \& Antoniadis(2016)]{2016ApJ...833L..12V} van Staden, A.~D., \& Antoniadis, J.\ 2016, \apjl, 833, L12 
\bibitem[Vigeland et al.(2018)]{2018ApJ...855..122V} Vigeland, S.~J., Deller, A.~T., Kaplan, D.~L., et al.\ 2018, \apj, 855, 122 
\bibitem[Wadiasingh et al.(2017)]{2017ApJ...839...80W} Wadiasingh, Z., Harding, A.~K., Venter, C., B{\"o}ttcher, M., \& Baring, M.~G.\ 2017, \apj, 839, 80 
\bibitem[Yao et al.(2017)]{2017ApJ...835...29Y} Yao, J.~M., Manchester, R.~N., \& Wang, N.\ 2017, \apj, 835, 29 
\bibitem[Zahn(1977)]{1977A&A....57..383Z} Zahn, J.-P.\ 1977, \aap, 57, 383 


\end{thebibliography}
\end{document}